\documentclass[fleqn,usenatbib]{mnras}

\usepackage{newtxtext,newtxmath}

\usepackage[T1]{fontenc}

\usepackage{graphicx}	
\usepackage{amsmath}	
\usepackage{mathtools}
\usepackage{multirow}
\usepackage{arydshln}
\usepackage{ulem}
\usepackage{booktabs}

\renewcommand{\vec}[1]{\mathbfit{#1}}
\newcommand{\mat}[1]{\mathbfss{#1}}

\newcommand\T{\rule{0pt}{2.6ex}}       
\newcommand\B{\rule[-1.2ex]{0pt}{0pt}} 

\title[Horndeski Gravity with Phantom Crossing]{Constraints on Horndeski gravity with phantom crossing}

\author[K. Naidoo, J. Hallam, T. Baker and S. Sirera]{
Krishna Naidoo,$^{1,2}$\thanks{E-mail: \href{mailto:krishna.naidoo@port.ac.uk}{krishna.naidoo@port.ac.uk}}
James Hallam,$^{1}$
Tessa Baker$^{1}$
and Sergi Sirera$^{1}$
\\
$^{1}$Institute of Cosmology and Gravitation, University of Portsmouth, Burnaby Road, Portsmouth, PO1 3FX, UK\\
$^{2}$Max-Planck-Institut f\"ur Astronomie, K\"onigstuhl 17, D-69117 Heidelberg, Germany
}

\date{Accepted XXX. Received YYY; in original form ZZZ}

\pubyear{\the\year{}}

\begin{document}
\label{firstpage}
\pagerange{\pageref{firstpage}--\pageref{lastpage}}
\maketitle

\begin{abstract}
Gravity models in which the dark energy equation of state crosses $w=-1$, also known as the phantom divide, have received extensive interest due to recent analyses favouring this behaviour. We introduce a new subclass of Horndeski scalar-tensor models capable of generating phantom crossing, whilst remaining minimally coupled to matter: the Asymptotic Cubic Galileon (ACG) models. We show that ACG models can jointly fit the expansion history inferred from observations of the Planck cosmic microwave background, baryon acoustic oscillation measurements from the Dark Energy Spectroscopic Instrument, and distance-ladder supernovae measurements from the Dark Energy Survey. We then demonstrate that requiring a positive galaxy-ISW cross-correlation, together with physically viable screening behaviour, substantially restricts the viable ACG parameter space.
Model comparison metrics, including $\chi^{2}$ and Bayesian evidence, favour both ACG and $w_{0}w_{a}$CDM models over $\Lambda$CDM, with ACG providing a fit of comparable quality to $w_{0}w_{a}$CDM. Crucially, ACG models ground the observationally preferred $w_{0}w_{a}$CDM behaviour in a robust Lagrangian formulation. This enables interpretation beyond mere phenomenological fits, and motivates further tests of these models on nonlinear scales.
\end{abstract}




\section{Introduction}

The standard model of cosmology, dominated by cold dark matter (CDM) and dark energy (DE), has underpinned precision cosmology for over two decades. Despite its many successes, the model relies on the existence of CDM and DE in the form of a cosmological constant, $\Lambda$, but neither of these have been directly observed. Moreover, there remain persistent theoretical challenges in reconciling both CDM and $\Lambda$ with the standard model of particle physics.

Observations of the Universe's expansion history have placed intense scrutiny on the nature of DE. Measurements of baryon acoustic oscillations (BAO) from the Dark Energy Spectroscopic Instrument \citep[DESI;][]{DESI_BAO_DR2}, along with Type Ia supernova data from Union3 \citep{Union3} and the Dark Energy Survey \citep[DES;][]{DES_SN_Y5}, indicate a preference for dynamical DE -- models in which the equation-of-state (EoS) of DE evolves over time. In particular, current data favor scenarios where DE exhibits phantom behavior at early times (i.e. an EoS less than $-1$), before crossing the phantom divide at a redshift of $z \approx 0.5$. Notably, these conclusions appear largely insensitive to the specific parametrization of DE's EoS \citep{Lodha2025}.

Phenomenological models of DE, such as the CPL parameterization \citep{ChevallierPolarski2001,Linder2003}, allow us to explore a vast set of behaviours efficiently. However, they lack the power to explain and predict other physical phenomena. If we can couch an evolving DE EoS within a Lagrangian formulation of gravity, we can make predictions for the impact of the model upon other observables, e.g. galaxy clustering, growth rates and weak lensing. These in turn allow the gravity model to be validated or eliminated by other data sets.

To achieve this Lagrangian-level realization, we turn to Horndeski gravity. Horndeski gravity \citep{Horndeski1974} is the most general class of scalar–tensor theories in four dimensions that yield second-order field equations for both the metric and the scalar field. Its broad theoretical framework encompasses a wide range of modified gravity models, including Dvali–Gabadadze–Porrati \citep[DGP;][]{DGP2000}, $f(R)$ \citep{HuSawicki2007}, and Galileon models \citep{Nicolis2009}. This versatility makes Horndeski theory particularly well suited for exploring extensions beyond general relativity, as well as scalar-field-based descriptions of DE. 

Horndeski theories with non-minimal coupling have been shown to be capable of producing the phantom crossing behaviour preferred by observations \citep{Ye2025,Wolf2025,Wolf2025b}. However, these models often predict significant modifications to the growth of structure, potentially placing them in tension with large-scale structure observations and galaxy–Integrated Sachs–Wolfe \citep[ISW;][]{SachsWolfe1967} cross-correlation measurements. This can arise from significant gradients in the modification to the lensing potential \citep{Renk2017, Seraille2024}. For minimally coupled kinetic braiding models \citep[in contrast to pure k-essence, where stable phantom crossing is forbidden;][]{Vikman:2004dc}, \citet{Tsujikawa2025} and \citet{Wolf2026} show how to cross the phantom divide by breaking shift symmetry with non-trivial potentials in the scalar field. \citet{Linder2025} explored the unusual kinetic structure required when attempting to realise phantom-crossing behaviour in shift-symmetric cubic Horndeski models, and \citet{Calderon2026} subsequently clarified that purely shift-symmetric Horndeski models without a potential do not realise the desired stable phantom crossing; adding a linear potential enables crossing but is difficult to reconcile with current observations. Complementary to these Lagrangian-level constructions, \citet{Cataneo2026} and \cite{Thummel2026} recently performed a non-parametric exploration of minimally coupled, luminal kinetic gravity braiding models using a stable EFT basis, identifying viable phantom-crossing models consistent with current background and linear large-scale-structure probes.

In this paper, we identify a class of models capable of explaining the current preference for dynamical DE and accommodating phantom crossing, whilst maintaining consistency with a positive ISW-galaxy cross-correlation signal. The model closely follows the Cubic Galileon model but with shift symmetry broken by adding multiplicative $\phi$-dependent terms either to the kinetic or kinetic-braiding terms, instead of changing the potential \citep[as carried out in][]{Tsujikawa2025, Wolf2026}. In doing so, we lay the groundwork for constraining viable models using linear perturbation theory and in the non-linear regime, using numerical simulations.

Our analysis is based on the flexible \texttt{Hi-COLA}\footnote{\href{https://github.com/Hi-COLACode/Hi-COLA}{https://github.com/Hi-COLACode/Hi-COLA}} software package \citep{Wright2023} that solves the background evolution of the scalar field and expansion history for any user defined Horndeski model. These background quantities can be passed onto a COmoving Lagrangian Acceleration \citep[COLA;][]{Tassev2013} $N$-body simulator within \texttt{Hi-COLA}, using a modified version of \texttt{COLASolver} \citep{Winther2023}, to construct cosmological simulations of large scale-structure in Horndeski gravity.

The paper is organised as follows. In Section~\ref{theory}, we describe the cosmological models evaluated in this paper. In Section~\ref{sec_expansion_and_growth}, we outline the expressions governing the background evolution of these models, and the scalar field evolution in Horndeski gravity, the EoS, linear perturbations theory and ISW effects. In Section~\ref{observations}, we outline the observational data used to constrain the expansion history and describe the Markov Chain Monte Carlo (MCMC) methods employed to obtain constraints on both cosmological and Horndeski-specific parameters. In Section~\ref{results}, we analyse the constraints on a range of Horndeski models, before discussing the broader implications for viable scalar-tensor theories in Section~\ref{discussion}. Finally, in Section~\ref{conclusions}, we summarise our findings and present the main conclusions of this work.

\section{Cosmological Models}
\label{theory}

In this section, we describe the theoretical models explored in this paper. We begin by describing the phenomenological dynamical DE model, using CPL parameterisation, and then we describe Horndeski gravity and the specific ACG models analysed in this paper. 

In \citet{Wright2023} densities of the constituents of matter, radiation, dark energy and the scalar field $\phi$, were defined in terms of the the dimensionless $\Omega$ functions. In this paper, we switch to the dimensionfull densities $\rho$. $\Omega$ and $\rho$ are related, for a constituent $i$, by the expression
\begin{equation}
    \Omega_{i} = \frac{8\pi G}{3 H^{2}}\rho_{i},
\end{equation}
where $G$ is Newton's gravitational constant and $H$ is the Hubble expansion rate. The switch from $\Omega$ to $\rho$ allows us to determine the density of any given constituent without needing the current expansion rate, simplifying calculations made in the rest of the paper. Furthermore, we can absorb constants in the above equation to define densities in terms of
\begin{equation}
    \hat{\rho}_{i}=\frac{8\pi G}{3 H_{0}^{2}}\rho_{i},
\end{equation}
using the expression $H=H_{0}E$ for the Hubble expansion, where $H_{0}$ is the Hubble constant.

\subsection{Dynamical Dark Energy}

In dynamical DE models, the DE density is no longer a constant but exhibits a time-varying energy density following the general form
\begin{equation}
    \hat{\rho}_{\mathrm{DE}}(a) = \Omega_{\mathrm{DE},0}\exp\left(-3\int_{1}^{a}\frac{1+w_{\mathrm{DE}}(a')}{a'}\mathrm{d}a'\right),
    \label{eq_DE_eos}
\end{equation}
where $\Omega_{\mathrm{DE}}$ is the fraction of DE and the subscript $0$ used to denote the fraction now (i.e. at redshift $z=0$), and $w_{\mathrm{DE}}(a)$ is the time-varying DE EoS. The Hubble expansion rate for dynamical DE is expressed by replacing $\hat{\rho}_{\Lambda}$ with $\hat{\rho}_{\mathrm{DE}}$. If we assume the EoS is characterised by the CPL \citep{ChevallierPolarski2001, Linder2003} parameterisation
\begin{equation}
    w_{\mathrm{DE}}(a) = w_{0} + w_{a}(1-a),
\end{equation}
where $w_{0}$ is the EoS today and $w_{a}$ defines a gradient, the energy density given by Eq.~\ref{eq_DE_eos} simplifies to
\begin{equation}
    \hat{\rho}_{\mathrm{DE}} = \Omega_{\mathrm{DE},0}\,a^{-3(1+w_{0}+w_{a})}e^{3w_{a}(a-1)}.
\end{equation}
We will refer to models of dynamical DE, following the above CPL parameterisation, as $w_{0}w_{a}$CDM.

\subsection{Horndeski Gravity}
\label{sec_Horndeski_theory}

As noted previously, the framework of Horndeski gravity allows us to perform generic computations for any (second-order) metric theory of gravity coupled to a single scalar field \citep{Horndeski1974}. We restrict our analysis to the luminal subset of Horndeski gravity, strictly enforcing the speed of gravitational waves to equal the speed of light \citep{Creminelli:2017sry,Ezquiaga:2017ekz,Baker2017}. While applying the stringent bound from the GW170817 standard siren event to cosmological scales is non-trivial, given the detection frequency roughly matches the typical effective field theory cutoff scale of DE models \citep{deRham:2018red}, we adopt the conservative approach of requiring exact luminality. For further discussions, see e.g. \cite{Harry:2022zey,Baker:2022rhh,Baker:2022eiz,Sirera:2023pbs,Atkins:2024nvl}.
 
The luminal Horndeski Lagrangian is defined in terms of the scalar field $\tilde\phi$ and its kinetic term $\tilde{X}=-\frac{1}{2}\nabla^{\mu}\tilde{\phi}\nabla_{\mu}\tilde{\phi}$. The action contains three unspecified functions of $\tilde{\phi}$ and $\tilde{X}$, which can be tuned to recover specific gravity models: a kinetic function $\tilde{K}(\tilde{\phi},\tilde{X})$, a braiding function $\tilde{G}_{3}(\tilde{\phi},\tilde{X})$, and a gravitational coupling function $\tilde{G}_{4}(\tilde{\phi})$:
\begin{equation}
    S =\int \mathrm{d}^{4}x\,\sqrt{-g}\,\Big(\tilde{G}_{4}(\tilde{\phi})R+\tilde{K}(\tilde{\phi},\tilde{X})-\tilde{G}_{3}(\tilde{\phi}, \tilde{X})\Box\tilde{\phi}
    -M_{\mathrm{P}}^{2}\Lambda + \mathcal{L}_{m}\Big)
    \label{eq:luminal_horndeski_action}
\end{equation}
where $\mathcal{L}_{m}$ is the matter Lagrangian and $M_{P}^{2}=1/(8\pi G)$ the Planck mass. Tilde symbols delineate dimensionful quantities. In what follows, we switch to working in terms of non-tilde dimensionless quantities, obtained by normalising the tilde quantities by appropriate mass and time scales. 
For example, $\phi=\tilde{\phi}/M_s$ where $M_s$ is a mass scale associated to the scalar field. The procedure for normalising the Horndeski functions $\tilde{K}, \tilde{G}_3$ and $\tilde{G}_4$ is given in Appendix \ref{subsec:mass_scales} \citep[see also Appendix A of][]{Wright2023}.

We include in Eq.~\ref{eq:luminal_horndeski_action} a cosmological constant $\Lambda$, which can be interpreted as either a true cosmological constant or as the minimum of a scalar field potential. For the latter interpretation, the minima is often absorbed into the definition of kinetic term $\tilde{K}$; for example, \cite{Wolf2026} considered Horndeski models where the function $\tilde{K}$ contains a quadratic potential term.\footnote{With the form
$\tilde{K}\sim \frac{1}{2}\tilde{X} - \tilde{V}(\tilde{\phi}),$ where $\tilde{V}=\tilde{V}_0+ \frac{1}{2}m^2\tilde{\phi}^2$ and $V_0$ acts like a cosmological constant.} Our definitions will always absorb any such constant in the $\Lambda$ term of Eq.~\ref{eq:luminal_horndeski_action}. This enables us to quantify the departure from $\Lambda$CDM by defining the fraction of DE provided by the scalar field
\begin{equation}
    f_{\phi}(a)=\frac{\Omega_{\phi}(a)}{\Omega_{\Lambda}(a)+\Omega_{\phi}(a)},
    \label{eq_f_phi}
\end{equation}
where $\Omega_{\phi}$ is the fractional density of the scalar field. This gives the relative strength of the scalar field contribution at a specific scale factor $a$ with respect to a cosmological constant-like component. With this definition any luminal Horndeski model is equivalent to $\Lambda$CDM when $f_{\phi}\rightarrow0$ for all values of a. In the next subsections, we will describe the models considered in this paper, which are summarised for reference in Table~\ref{tab_horndeski_models}.

\begin{table}
    \centering
    \begin{tabular}{lll}
        \hline\hline
        Model & $K(\phi,X)$ & $G_{3}(\phi,X)$\\
        \hline
        Cubic Galileon & $k_{1}X$ & $g_{31}X$\\
        ESS & $k_{1}X+k_{2}X^{2}$ & $g_{31}X+g_{32}X^{2}$\\
        \midrule
        \textbf{Asym. Cubic Galileon} & $k_{1}\mathcal{K}(\phi)X$ & $g_{31}\mathcal{G}(\phi)X$\\
        \cmidrule(lr){1-3}
        \quad Growing $\mathcal{G}(\phi)$ & $k_{1}X$ & $g_{31}\left(1+c_{g_{3}}\phi\right)X$\\
        \quad Decaying $\mathcal{K}(\phi)$& $k_{1}\exp\left(-c_{k}\phi\right)\,X$ & $g_{31}X$\\
        \hline\hline
    \end{tabular}
    \caption{Summary of the minimally coupled Horndeski models ($G_4=1/2$) considered in this paper. The upper section contains the shift-symmetric Cubic Galileon and Extended Shift-Symmetric (ESS) models. The lower section introduces the general form of Asymptotic Cubic Galileon (ACG) models, characterised by a phi-dependent kinetic function $\mathcal{K}(\phi)$ and/or braiding function $\mathcal{G}(\phi)$. The two rows beneath the ACG entry correspond to the specific Growing $\mathcal{G}(\phi)$ and Decaying $\mathcal{K}(\phi)$ models analysed in this paper.}
    \label{tab_horndeski_models}
\end{table}

\subsection{The Asymptotic Cubic Galileon Model}
\label{sec:cubic_galileon}

Within the Horndeski gravity framework, the Cubic Galileon (CG) model \citep{Nicolis2009,Deffayet:2009wt, C_dric_Deffayet_2010} is one of the simplest and most extensively studied models in the field. The model is expressed with the following Horndeski free functions
\begin{equation}
    K(X) = k_{1}X, \quad G_{3}(X) = g_{31}X, \quad G_{4} = \frac{1}{2},
    \label{eq:cg_functions}
\end{equation}
where $k_{1}$ and $g_{31}$ are constants. The model is insensitive to the absolute value of the scalar field $\phi$, since the model is expressed purely in terms of $X$ and therefore only sensitive to the derivative of the scalar field $\phi^{\prime}$. Models expressed solely in term of $X$ are referred to as shift symmetric, meaning the scalar field $\phi$ can be shifted by an arbitrary constant without affecting the dynamics of the model. 
Shift symmetric models preserve the shift-charge density \citep{Bellini2014}, which results in an attractor solution to the scalar field evolution known as the Tracker ansatz. The attractor behaviour leads to a reduction of free parameters, by converting some of the equations of motion into constraint equations.
For the CG the Tracker ansatz reduces the number of free parameters from two ($k_{1}$ and $g_{31}$) to one, which is given by the quantity $f_{\phi}$ (at redshift z=0),
\begin{equation}
   k_{1} = -6\,f_{\phi}(a=1)\,\Omega_{\mathrm{DE},0}\quad\text{and}\quad g_{31} = 2 \,f_{\phi}(a=1)\,\Omega_{\mathrm{DE},0}\,.
\end{equation}

Studies constraining the CG with the cosmic microwave background (CMB) from Planck \citep{Barreira2014} fix $f_{\phi}=1$, meaning the scalar field is the sole contributor to DE and the model has no additional free parameters in comparison to $\Lambda$CDM. This removes the explicit cosmological constant contribution but does not eliminate the need for parameter or initial-condition tuning to reproduce the observed late-time acceleration. These CG models also predict negative large-scale structure (LSS) to CMB cross-correlations (see Sec.~\ref{sec_ISW_theory}), placing them in significant tension with the positive cross-correlations measured from observations \citep{Renk2017}. To ensure positive LSS-CMB cross-correlations, viable CG models must have $f_{\phi} < 1$, meaning they must have a significant contribution from the cosmological constant or scalar field minimum potential.

Even with $f_\phi$ free, CG models cannot produce an EoS that crosses the phantom divide (see Appendix~\ref{app_shift_symmetric}; \citealt{Hallam2026}). More generally, stable phantom crossing is highly restricted in shift-symmetric Horndeski models \citep{Traykova2021,Tsujikawa2025,Linder2025}, and purely shift-symmetric models without a potential do not realise the desired stable crossing behaviour \citep{Calderon2026}. These models are therefore unlikely to be viable given the current preference for phantom crossing \citep{DESI_BAO_DR2}.

Based on these considerations, we introduce here a new class of Asymptotic Cubic Galileon (ACG) gravity models. ACG models are designed to follow the CG behaviour at early times, producing the early phantom behaviour preferred by observations, but diverging at late times, allowing the model to cross the phantom divide. The theoretical framework underlying ACG models, including phantom-crossing behaviour, is discussed in detail in the companion paper \citep{Hallam2026}. We find that phantom crossing in ACG can be achieved by breaking shift symmetry and introducing a $\phi$-dependence to either the Kinetic $K$ term or the kinetic-braiding term $G_{3}$. For phantom crossing to occur, we require either the kinetic term $K$ to decrease, dampening towards but not equaling zero, or the kinetic-braiding term $G_{3}$ to increase. Although both mechanisms can induce phantom crossing, they do so through distinct modifications to the scalar-field dynamics.

We define ACG models with the following functional form
\begin{equation}
\begin{split}
    K(\phi,X)=k_{1}\mathcal{K}(\phi)X,\quad G_{3}(\phi,X)=g_{31}\mathcal{G}(\phi)X,\quad G_{4}=\frac{1}{2}.
    \label{eq:ACG_KG3}
\end{split}
\end{equation}
To ensure ACG follows CG models at early times we require that $\mathcal{K}(\phi(a\rightarrow 0))\rightarrow 1$ and similarly, $\mathcal{G}(\phi(a\rightarrow 0))\rightarrow 1$.

This can be achieved using the smallness of the scalar field at early times, $\phi(a\rightarrow 0)\rightarrow 0$, allowing $\mathcal{K}(\phi)$ and $\mathcal{G}(\phi)$ to become non-trivial once $\phi$ grows. The direction of this evolution is physically motivated by the phenomenological requirements of Rolling Galileon models. In particular, \citet{Hallam2026} show that viable models favour an increasing braiding strength relative to the kinetic sector, which can be realised either through a growing $\mathcal{G}(\phi)$ or a decaying $\mathcal{K}(\phi)$. Conversely, a decaying braiding term or growing kinetic term drives the theory in the opposite direction. With this in mind, we consider the following specific ACG models:
\begin{itemize}
\item \textbf{Growing $\mathcal{G}(\phi)$}: with a linearly growing kinetic-braiding term
\begin{equation}
\label{Kdef}
\mathcal{K}(\phi) = 1,\quad\quad \mathcal{G}(\phi) = 1+c_{g_{3}}\phi.
\end{equation}
\item \textbf{Decaying $\mathcal{K}(\phi)$}: with an exponentially decaying kinetic term
\begin{equation}
\label{Gdef}
\mathcal{K}(\phi) = \exp\left(-c_{k}\phi,\right),\quad \mathcal{G}(\phi) = 1.
\end{equation}
\end{itemize}
The particular linear and exponential forms adopted here are simple phenomenological parameterisations rather than unique theoretically preferred functions. Their common physical behaviour---an increasing braiding strength relative to the kinetic sector---is more general, and is explored in detail by \citet{Hallam2026}. An advantage of the ACG assumption is that it fixes the initial scalar-field velocity, $\phi'$, leaving each model with only one additional free parameter relative to the CG model and two relative to $\Lambda$CDM, the same number of additional free parameters as dynamical DE. Here we consider the Growing $\mathcal{G}(\phi)$ and Decaying $\mathcal{K}(\phi)$ parameterisations separately, while \citet{Hallam2026} also consider models combining both behaviours.

Although the ACG models, and indeed the observationally preferred $w_{0}w_{a}$CDM model, exhibit a DE EoS satisfying $w_{\rm DE}<-1$, this does not imply a violation of the null energy condition (NEC) by the total cosmic fluid. Throughout the evolution considered here, the effective equation of state of the Universe remains $w_{\rm eff}>-1$, since the phantom phase occurs predominantly at $z\gtrsim0.5$, when matter still contributes significantly to the total energy density. During the epoch of dark-energy domination, both models satisfy $w_{\rm DE}>-1$, and hence the total cosmic fluid never violates the NEC. Furthermore, although phantom behaviour can be associated with ghost or gradient instabilities in simpler scalar-field theories such as pure k-essence, this is not generally the case for kinetic gravity braiding theories. Accordingly, all viable models considered here are additionally required to satisfy the no-ghost and gradient-stability conditions throughout their evolution.

\section{Expansion and Growth of Structure}
\label{sec_expansion_and_growth}

In this section we describe the functions governing the background expansion and evolution of the scalar field, the effective EoS, growth of structure and the ISW effect in Horndeski gravity. We will provide the full expressions for the reduced Horndeski framework, into which we then substitute the models in Eqs.~\ref{Kdef} and \ref{Gdef}.

\subsection{Friedmann Equation}

The first Friedmann equation, defining the expansion rate of the Universe, can be expressed generally as 
\begin{equation}
    H^{2}(a) = H_{0}^{2}\Big[\hat{\rho}_{\gamma}(a)+\hat{\rho}_{\mathrm{b}}(a)+\hat{\rho}_{\mathrm{c}}(a)+\hat{\rho}_{\nu}(a)+\hat{\rho}_{\Lambda}+\hat{\rho}_{\phi}(a)\Big].
    \label{eq_hubble_expansion}
\end{equation}
The density of constituent $i$ is given b $\hat{\rho}_{i}$ where $\gamma$ denotes photons, $b$ baryons, $c$ CDM, $\nu$  neutrinos, $\Lambda$ for the cosmological constant and $\phi$ for a scalar field. Hence, $\Lambda$CDM is recovered when $\hat{\rho}_{\phi}=0$. The photon density is expressed in terms of the CMB temperature today $T_{\gamma,0}=2.7255\,\textrm{K}$ as 
\begin{equation}
    \hat{\rho}_{\gamma}(a)=\frac{C_{\gamma}}{h^{2}a^{4}},\quad C_{\gamma}=\frac{8\pi^{3}GT_{\gamma,0}^{4}}{45}\approx 2.473012\times10^{-5}
\end{equation}
where $h=H_{0}/100$ is the normalised Hubble constant. The baryon and CDM density are given by $\hat{\rho}_{\mathrm{b}}=\Omega_{\mathrm{b},0}/a^{3}$ and $\hat{\rho}_{\mathrm{c}}=\Omega_{\mathrm{c},0}/a^{3}$, respectively, where $\Omega_{\mathrm{b},0}$ and $\Omega_{\mathrm{c},0}$ are the fractional energy density of baryons and CDM in the Universe today. The density of the cosmological constant is $\hat{\rho}_{\Lambda}=\Omega_{\Lambda,0}$ where $\Omega_{\Lambda,0}$ is the fractional density of DE today -- which is computed by assuming zero curvature. 

\subsection{Massive Neutrinos}
\label{sec_mass_neu}

The neutrino density is evaluated as two components: massless relativistic species (denoted with an $\mathrm{UR}$) and massive non-relativistic species (denoted with a $\mathrm{NR}$), 
\begin{equation}
    \hat{\rho}_{\nu}(a) = \hat{\rho}_{\nu}^{\mathrm{UR}}(a)+\hat{\rho}_{\nu}^{\mathrm{NR}}(a).
\end{equation}
The relativistic density simply scales as
\begin{equation}
    \hat{\rho}_{\nu}^{\mathrm{UR}}(a)=\frac{N_{\mathrm{UR}}}{3}N_{\mathrm{eff}}\left(\frac{7}{8}\right)\left(\frac{4}{11}\right)^{4/3}\hat{\rho}_{\gamma},
\end{equation}
where $N_{\mathrm{eff}}$ is the effective number of neutrino species, and $N_{\mathrm{UR}}$ the number of massless neutrino species, with an implicit assumption that there can be no more than three in total. The density of non-relativistic neutrinos is given by
\begin{equation}
    \hat{\rho}_{\nu}^{\mathrm{NR}}(a)=\frac{C_{\nu}}{h^{2}a^{4}}\sum_{i=1}^{N_{\nu}}\int_{0}^{\infty}\frac{x^{2}\sqrt{x^{2}+y^{2}}}{e^{x}+1}\mathrm{d}x, \quad \text{where}\;\; y=\frac{am_{\nu,i}}{T_{\nu,0}},
\end{equation}
$N_{\nu}$ the number of massive neutrino species, $m_{\nu,i}$ the mass of neutrino species $i$ \citep{Elbers2025}, $T_{\nu,0}$ the temperature of relic neutrinos from the Big Bang and $C_{\nu}=8GT_{\nu,0}^{4}/(300\pi)$. Since we know neutrino mass must be positive \citep{Ahmad2001,Fukuda1998} we can use neutrino mass as a diagnostic tool to test whether a model fits observations correctly. Within $\Lambda$CDM, observations currently prefer a negative neutrino mass \citep{Craig2024,ElbersDESI2025}, a result which is alleviated in dynamical DE models \citep{ElbersDESI2025} and driven by a phantom EoS for DE at early times \citep{Yang2026}. Following \citet{Elbers2025}, we allow neutrinos to have an effective negative mass by replacing the above equation with the absolute mass and switching the sign of the above equation in the presence of negative neutrino mass.

\subsection{Scalar Field Evolution}

The expansion rate and scalar field equations are governed by a set of coupled ordinary differential equations \citep[ODE;][]{Kimura2012, Bellini2014, Sirera2026}. Here, we follow the definitions used in the \texttt{Hi-COLA} code \citep{Wright2023}, a lightweight simulation package for Horndeski gravity. Further technical details on the \texttt{Hi-COLA} solver can be found in Appendix \ref{solver}.

The density of the scalar field is given by
\begin{equation}
    \begin{split}
        \hat{\rho}_{\phi}=&\left(\frac{1}{2G_{4}}-1\right)\sum_{i}\hat{\rho}_{i}+ \frac{1}{3G_{4}}\Bigg(X K_{X} - \frac{K}{2}+ 3E^{2}X\phi^{\prime}G_{3X}\\
        &-XG_{3\phi}-3E^{2}\phi^{\prime}G_{4\phi}\Bigg),
    \end{split}
\end{equation}
where we absorb and remove mass scaling terms for brevity and clarity. Derivatives with respect to $\phi$ or $X$ are denoted as subscripts, meaning a term $F_{X}=\partial_{X}F$ is a derivative with respect to $X$. Lastly, derivatives with respect to time, denoted with a dot in \citep{Kimura2012, Bellini2014, Sirera2026}, have been redefined with derivatives with respect to $x=\log a$ denoted with a prime $^{\prime}$. In this definition, 
\begin{equation}
    X=\frac{E^{2}\phi^{\prime2}}{2},
\end{equation}
which expresses sensitivity to the derivative of the scalar field.

The second Friedmann (Raychaudhuri) equation can be rearranged to give an explicit expression for the evolution of the normalised expansion rate,
\begin{equation}
    \label{eq:E_prime}
    E^{\prime} = \frac{\mathcal{A}}{\mathcal{B}}E
\end{equation}
where the auxiliary functions $\mathcal{A}$ and $\mathcal{B}$ are given in Appendix~\ref{app_background}. Similarly, the scalar-field equation of motion can be written as
\begin{equation}
    \label{eq:phi_primeprime}
    \phi^{\prime\prime}=-\left(\frac{B_{1}}{A}+\phi^{\prime}\right)\frac{E^{\prime}}{E}-\frac{B_{2}}{A}.
\end{equation}
where $A$, $B_1$, and $B_2$ are defined in Appendix~\ref{app_background}.

In \texttt{Hi-COLA}, the variables $\phi$, $\phi^{\prime}$, and $E$ are evolved simultaneously. The field value $\phi$ is updated using $\phi^{\prime}$, while $\phi^{\prime}$ is evolved using Eq.~\ref{eq:phi_primeprime}. The expansion rate $E$ is obtained from the closure relation, Eq.~\ref{eq_hubble_expansion}, with Eq.~\ref{eq:E_prime} providing an initial guess for the iterative solution. Further details of the numerical implementation are provided in Appendix~\ref{solver}.

\subsection{Effective Equation-of-State}
\label{sec_EoS_theory}

Following  \citep{Bellini2014}, we can define an effective EoS for the scalar field in Horndeski gravity using the expression
\begin{equation}
    w_{\phi}=\frac{\tilde{\mathcal{P}}}{\tilde{\mathcal{E}}}
\end{equation}
where
\begin{equation}
    \tilde{\mathcal{E}}=\frac{H_{0}^{2}}{2G_{4}}\Big[-K+2XK_{X}-2XG_{3\phi}+6E^{2}\phi^{\prime}\left(XG_{3X}-G_{4\phi}\right)\Big],
\end{equation}
\begin{equation}
\begin{split}
    \tilde{\mathcal{P}}=\frac{H_{0}^{2}}{2G_{4}}\Bigg[K-2X\big(&G_{3\phi}-2G_{4\phi\phi}\big)+4E^{2}\phi^{\prime}G_{4\phi}\\
    &-2E^{2}\left(\frac{E^{\prime}}{E}\phi^{\prime}+\phi^{\prime\prime}\right)\left(XG_{3X}-G_{4\phi}\right)\Bigg].
\end{split}
\end{equation}

The combined EoS for the DE sector is given by
\begin{equation}
    w_{\mathrm{DE}}=\frac{\hat{P}_{\Lambda}+\hat{P}_{\phi}}{\hat{\rho}_{\Lambda}+\hat{\rho}_{\phi}}=\frac{-\hat{\rho}_{\Lambda}+w_{\phi}\hat{\rho}_{\phi}}{\hat{\rho}_{\Lambda}+\hat{\rho}_{\phi}}\equiv\frac{-\Omega_{\Lambda}+w_{\phi}\Omega_{\phi}}{\Omega_{\Lambda}+\Omega_{\phi}}
\end{equation}
where $\Lambda$ denotes the contribution to the total DE from the cosmological constant. See \citet{Hallam2026} for a detailed discussion of what drives phantom crossing in ACG models.

\subsection{Linear Perturbation Theory}

The perturbed metric can be expressed as
\begin{equation}
    ds^2=-(1+2\Phi)dt^2+a^2(1-2\Psi)d\vec{x}^2,
\end{equation}
where $\Phi$ and $\Psi$ denote the scalar gravitational potentials in the Newtonian gauge, corresponding respectively to perturbations of the temporal and spatial components of the Friedmann–Lema\^{i}tre–Robertson–Walker (FLRW) metric.

We can define the modified Poisson equation as
\begin{equation}
    \nabla^{2}\Psi(\vec{x},a) = \frac{3}{2}H_{0}^{2}\frac{\Omega_{m,0}}{a}\mu(a)\delta(\vec{x},a),
\end{equation}
where $\vec{x}$ is a position in space and $\mu$ defines a modification to the strength of gravity. In general, $\mu$ is a scale- and time-dependent function. However, on scales well inside the cosmological horizon and below the Compton wavelength of the scalar degree of freedom, the quasistatic approximation can be applied, provided $c_{s}^{2}$ is not too close to zero \citep{Pogosian2016}. This allows time derivatives of the perturbations to be neglected relative to spatial gradients. We further assume that the scalar sector does not introduce an additional effective mass scale within the range of scales considered, so that any residual scale dependence can be neglected \citep[see, e.g.,][]{Baker2014}. Under these assumptions, the modifications become effectively scale-independent and the linear-theory solution can be written in terms of the $\alpha$-functions, $c_{s}^{2}$, and $D$, defined in Appendix~\ref{app_alphas}, as
\begin{equation}
    \mu_{L}(a) = \frac{1}{2G_{4}}\left[1+\frac{2}{c_{s}^{2}D}\left(\frac{\alpha_{B}}{2}+\alpha_{M}\right)^{2}\right],
\end{equation}
where the subscript $\mathrm{L}$ denotes the linear quasistatic solution.

Similarly, the Weyl-lensing potential is given by
\begin{equation}
    \nabla^{2}\left[\Psi(\vec{x},a)+\Phi(\vec{x},a)\right]=3H_{0}^{2}\frac{\Omega_{m,0}}{a}\Sigma(a)\,\delta(\vec{x},a),
    \label{eq_weyl}
\end{equation}
where $\Sigma$ defines a beyond-General Relativity (GR) modification to the lensing potential. As with $\mu_{L}$, the expression below is valid in the linear quasistatic limit,
\begin{equation}
    \Sigma_{L}(a)=\frac{1}{2G_{4}}\left[1+\frac{1}{c_{s}^{2}D}\left(\frac{\alpha_{B}}{2}+\alpha_{M}\right)\left(\alpha_{B}+\alpha_{M}\right)\right].
    \label{eq_Sigma_L}
\end{equation}

The linear growth function $D_{1}$ is given by the ODE
\begin{equation}
    D_{1}^{\prime\prime}=\frac{3}{2}\frac{\Omega_{m,0}}{E^{2}a^{3}}\mu_{L}D_{1}-\left(2+\frac{E^{\prime}}{E}\right)D_{1}^{\prime}.
\end{equation}
with $D_{1}(z=0)=1$ and the growth rate defined as
\begin{equation}
    f_{1}=\frac{\mathrm{d}\ln D_{1}}{\mathrm{d}\ln a}=(\ln D_{1})^{\prime}.
\end{equation}

\subsection{Integrated Sachs-Wolfe Effect}
\label{sec_ISW_theory}

The ISW \citep{SachsWolfe1967} is a second-order CMB anisotropy caused by the gain or loss of energy caused by photons traversing evolving gravitational potentials. This is effect is dominated by low redshift large angular modes well within the linear regime. The ISW is given by the integral
\begin{equation}
    \frac{\Delta T(\vec{n})}{T_{\gamma,0}} = \frac{1}{c^{3}}\int_{0}^{\chi_{\mathrm{LS}}}a^{2}H\,\frac{\mathrm{d}}{\mathrm{d}a}\bigg(\Psi(\chi\vec{n}, a) + \Phi(\chi\vec{n}, a) \bigg)\mathrm{d}\chi,
\end{equation}
where $\chi$ is the comoving distance, $\chi_{\mathrm{LS}}$ the comoving distance to last scattering, $c$ the speed of light and $\vec{n}$ is a unit vector pointing in a direction on the sky. We can express the lensing potential using Eq.~\ref{eq_weyl} as
\begin{equation}
    \Psi(\vec{x},a)+\Phi(\vec{x},a) = 3H_{0}^{2}\Omega_{m,0}\frac{\Sigma_{L}(a)D_{1}(a)}{a}\,\left[\nabla^{-2}\delta(\vec{x})\right]
\end{equation}
where we take the linear and quasistatic approximation 
\begin{equation}
    \delta(\vec{x},a)=D_{1}(a)\delta(\vec{x},a=1)=D_{1}(a)\delta(\vec{x}),
\end{equation}
allowing us to separate the spatial and temporal components of the density contrast. This factorisation follows from the scale-independent quasistatic limit discussed above, which assumes the Compton wavelength of the scalar field is comparable to or larger than the cosmological horizon. The temporal derivative of this function can be expressed as
\begin{equation}
\begin{split}
    \frac{\mathrm{d}}{\mathrm{d}a}\bigg(&\Psi(\vec{x},a)+\Phi(\vec{x},a)\bigg)=\\
    &3H_{0}^{2}\Omega_{m,0}\frac{\Sigma_{L}(a)D_{1}(a)}{a^{2}}\Big[f_{1}(a)+\zeta(a)-1\Big]\left[\nabla^{-2}\delta(\vec{x})\right],
\end{split}
\end{equation}
with
\begin{equation}
    \zeta(a) = \frac{\mathrm{d}\ln}{\mathrm{d} \ln a}\bigg(\frac{\Sigma_{\mathrm{L}}(a)}{\Sigma_{\mathrm{L}}(a=1)}\bigg).
    \label{eq_zeta}
\end{equation}
In the general relativistic limit, $\Sigma_{\mathrm{L}}(a)=1$ and therefore $\zeta(a)=0$, returning the familiar equation for the ISW in $\Lambda$CDM. The ISW can now be re-expressed as
\begin{equation}
\begin{split}
    &\frac{\Delta T(\vec{n})}{T_{\gamma,0}} = -\frac{3H_{0}^{3}\Omega_{m,0}}{c^{3}}\\
    &\times\int_{0}^{\chi_{\mathrm{LS}}}E(a)\Sigma_{L}(a)D_{1}(a)\,\Big[1-f_{1}(a)-\zeta(a)\Big]\,\Big[\nabla^{-2}\delta(\vec{x})\Big]\mathrm{d}\chi.
\end{split}
\end{equation}
Note, although the integral is a function of comoving distance $\chi$, $\chi$ can be mapped on to of the scale factor $a$. For convenience we will refer to these functions in terms of the scale factor $a$. 

Measurements of the ISW are given by cross-correlating the CMB with large scale structure. This is often measured with angular power spectra, defined generally
\begin{equation}
    C_{\ell}^{\mathrm{XY}} \approx \int \frac{W_{\mathrm{X}}(\chi) W_{\mathrm{Y}}(\chi)}{\chi^{2}}P^{\mathrm{XY}}(k_{\ell},a)\,\mathrm{d}\chi
\end{equation}
using the Limber approximation \citep{Loverde2008}, where $W_{\mathrm{X}}(\chi)$ is a window function for a source $\mathrm{X}$, $P^{\mathrm{XY}}$ is the cross-spectrum and 
\begin{equation}
    k_{\ell}=\frac{\ell+\frac{1}{2}}{\chi}.
\end{equation}
The CMB-galaxy cross-spectrum is given by
\begin{equation}
    C_{\ell}^{Tg} \approx \int \frac{W_{\mathrm{T}}(\chi)W_{\mathrm{g}}(\chi)}{\chi^{2}}D_{1}^{2}(a)\Sigma_{\mathrm{L}}(a)\frac{P(k_{\ell})}{k_{\ell}^{2}} \mathrm{d}\chi
\end{equation}
where we once again take the linear, quasistatic approximation to decouple the spatial and temporal components of the power spectra, a function of the Weyl potential (denoted with $\Psi+\Phi$) and the density fields (denoted with $\delta$)
\begin{equation}
    P^{\mathrm{Tg}}(k,a)=P^{\Psi+\Phi,\delta}(k,a)=D_{1}^{2}(a)\,\Sigma_{\mathrm{L}}(a)\,\frac{P(k)}{k^{2}}.
\end{equation}
The factor $k^{2}$ comes from solving the Weyl potential in Fourier space. The window function for the ISW is given by
\begin{equation}
    W_{\mathrm{T}}(a) = \frac{3H_{0}^{3}\Omega_{m,0}}{c^{3}}E(a)\Big[1-f_{1}(a)-\zeta(a)\Big]
\end{equation}
and the galaxy window function is given by
\begin{equation}
    W_{\mathrm{g}}(a) = b(a)\,n_{g}(a)
\end{equation}
assuming a simple linear bias $b$ function and normalised redshift selection function $n_{g}$. From this relation we can see that the cross-correlation simplifies to
\begin{equation}
\begin{split}
    C_{\ell}^{\mathrm{Tg}} \approx \frac{3 H_{0}^{3}\Omega_{m,0}}{c^{3}\left(\ell+\frac{1}{2}\right)^{2}}\int D_{1}^{2}(a)\,& \Sigma_{\mathrm{L}}(a)\,E(a)\,\Big[1-f_{1}(a)-\zeta(a)\Big]\,\\
    &\times \,b(a)\,n_{g}(a)\,P(k_{\ell})\,\mathrm{d}\chi.
\end{split}
\end{equation}
We can remove the product $b(a)n_{g}(a)P(k)$ from the integral to remove the sensitivity to the specifics of the galaxy survey and primordial power spectra. This results in an integral, that we call the ISW strength $S_{\mathrm{ISW}}$, that is solely reliant on the time-dependent growth functions $D_{1}$ and $f_{1}$, modifications to the lensing potential $\Sigma_{\mathrm{L}}$ and its logarithmic derivative $\zeta$, and $E$ the normalised Hubble expansion rate
\begin{equation}
    S_{\mathrm{ISW}}=\int D_{1}^{2}(a)\,\Sigma_{\mathrm{L}}(a)\,E(a)\,\Big[1-f_{1}(a)-\zeta(a)\Big]\,\mathrm{d}\chi.
    \label{eq_ISW_strength}
\end{equation}

\section{Observations}
\label{observations}

Constraints on model extensions to $\Lambda$CDM typically depend on modified Boltzmann solvers, such as \texttt{CAMB} \citep{Lewis2011} or \texttt{CLASS} \citep{Lesgourgues2011}, interfaced with a Markov Chain Monte Carlo (MCMC) sampler such as \texttt{Cobaya} \citep{Torrado2019}. However, the computation of the matter power spectrum in these solvers and CMB anisotropies can be prohibitively expensive, making sampling slow and parameter and model space exploration inefficient. In this paper, we construct a custom sampler, comparing predictions of the expansion rate and linear perturbations to observations without needing a full prediction of the matter power spectrum. This allows us to evaluate the model space of Horndeski gravity more effectively and establish models of interest given the current preference for dynamical DE. In this section we outline the observational measurements and likelihoods used to constrain the Horndeski models and explain the MCMC sampling methods used in this analysis.

\subsection{Observational Measurements and Likelihoods}
\label{sec_obs_like}

\subsubsection{Planck Compressed Likelihood}

To ensure our constraints on Horndeski gravity are consistent with Planck \citep{Planck2020}, we use a compressed form of the Planck likelihood following \citet{DESI_BAO_DR2} that compresses the information from Planck into three quantities: $\theta_{*}$ the angular scale of the first BAO peak, $w_{\mathrm{b}}=\Omega_{\mathrm{b},0}h^{2}$ the physical density of baryons and $w_{\mathrm{bc}}=\Omega_{\mathrm{m},0}h^{2}$ the physical density of baryons and CDM. The compressed form of the Planck likelihood only assumes that the early Universe follows $\Lambda$CDM, with the sensitivity to the late-time Universe marginalised \citep[see][]{Lemos2023}. The likelihood is given by
\begin{align}
    &\log\mathcal{L}_{\mathrm{P}} = -\frac{1}{2}\Big[\Delta \vec{d}^{T}\cdot\mat{C}_{\mathrm{P}}^{-1}\cdot\Delta\vec{d}+\log|\mat{C}_{\mathrm{P}}|+N_{\mathrm{P}}\log(2\pi)\Big],
    \label{eq_cmb_likelihood}\\
    &\Delta \vec{d}=\vec{d}-\vec{d}_{\mathrm{P}},
\end{align}
where the model prediction is given by
\begin{equation}
    \vec{d}=
    \begin{pmatrix}
        \theta_{*}\\ w_{\mathrm{b}}\\ w_{\mathrm{bc}}
    \end{pmatrix},\\
\end{equation}
and the Planck observations are given by the measurements 
\begin{equation}
    \vec{d}_{\mathrm{P}}=
    \begin{pmatrix}
        0.01041\\ 0.02223\\ 0.14208
    \end{pmatrix}
\end{equation}
and covariance
\begin{equation}
   \mat{C}_{\mathrm{P}}=10^{-9}\begin{pmatrix}
        0.006621 & 0.12444 & -1.1929\\
        0.12444 & 21.344 & -94.001\\
        -1.1929 & -94.001 & 1488.4
    \end{pmatrix},
\end{equation}
with $N_{\mathrm{P}}=3$ the length of the data vector. The scale of the acoustic peak in the CMB is given by 
\begin{equation}
    \theta_{*}=\frac{r_{*}}{\chi(z_{*})}
    \label{eq_theta_star}
\end{equation}
where $r_{*}$ is the BAO scale and $z_{*}$ the redshift at recombination, computed using the \citet{HuSugiyama1995} approximation. When evaluating the compressed Planck likelihood, the sampled value of $H_0$ enters through Eq.~\ref{eq_theta_star} and is thereby converted into the corresponding prediction for $\theta_{*}$. By comparing results from the Boltzmann solver \citep{Lewis2011} to the \citet{HuSugiyama1995} approximation for $z_{*}$, we find a small $w_{\mathrm{b}}$-dependent deviation which we remove by fitting and applying the following correction
\begin{equation}
    z_{*} = z_{*}^{\mathrm{HS}} + 91.95w_{\mathrm{b}}-3.9896,
\end{equation}
where $z_{*}^{\mathrm{HS}}$ is the \citet{HuSugiyama1995} approximation. Furthermore, we replace the prediction of $r_{*}$ of \citet{HuSugiyama1995} with an exact computation of  baryon and photon ratio
\begin{equation}
    R=\frac{3}{4}\frac{\rho_{\mathrm{b}}}{\rho_{\gamma}},
\end{equation}
that is used to compute $R$ at recombination and at matter-radiation equality. The approximation predicts $\theta_{*}$ with an accuracy such that the difference from the \texttt{CAMB} prediction is smaller than the observational measurement uncertainty.

Without these approximations the \citet{HuSugiyama1995} prediction provide values that are biased at such a level that they become incompatible with the predictions from \texttt{CAMB} and could bias our parameter constraints. We will denote constraints from the Planck compressed likelihood with the shorthand CMB.

\subsubsection{DESI BAO}

We use BAO measurements from Data Release 2 (DR2) of the DESI survey, including measurements from the Bright Galactic Survey (BGS), Luminous Red Galaxies (LRG), Emission Line Galaxies (ELG), Quasars (QSO) and Lyman-$\alpha$ (Ly$\alpha$) tracers. The survey is split into seven redshift bins with thirteen BAO measurements made in total. DESI measures the BAO as an angle on the sky $D_{\mathrm{M}}/r_{\mathrm{d}}$, as a line-of-sight measurement $D_{\mathrm{H}}/r_{\mathrm{d}}$ or as a volume average measurement $D_{\mathrm{V}}/r_{\mathrm{d}}$. 

For the theoretical predictions we follow \citet{DESI_BAO_DR2} in using the approximation
\begin{equation}
    r_{\mathrm{d}}\approx 147.05\,\left(\frac{w_{\mathrm{bc}}}{0.1432}\right)^{-0.23}\,\left(\frac{N_{\mathrm{eff}}}{3.04}\right)^{-0.1}\left(\frac{w_{\mathrm{b}}}{0.02236}\right)^{-0.13}\,\text{Mpc}
\end{equation}
for the sound horizon at the drag-epoch. In this work, we assume the number of effective relativistic species is $N_{\mathrm{eff}}=3.04$. The angular BAO is a ratio of the comoving distance, denoted here as $D_{\mathrm{M}}$ but equivalent to $\chi$ used previously, to the sound horizon at the drag-epoch $r_{\mathrm{d}}$. The line-of-sight BAO is a ratio $D_{\mathrm{H}}(z)=c/H(z)$ over $r_{\mathrm{d}}$, and the volume averaged BAO is given by
\begin{equation}
    D_{\mathrm{V}}(z)=\Big(zD_{\mathrm{M}}^{2}(z)D_{\mathrm{H}}(z)\Big)^{1/3}.
\end{equation}
The likelihood follows the same form as Eq.~\ref{eq_cmb_likelihood}, but with each redshift bin treated separately since the separate bins are not correlated. We will denote constraints from the DESI DR2 BAO with the shorthand BAO.

As a validation of our likelihood implementation, we verified that applying the same compressed CMB and DESI BAO likelihoods to $\Lambda$CDM and $w_0w_a$CDM reproduces the published DESI parameter constraints, finding consistent results.

\subsubsection{DES Supernovae}

We use the DES-Dovekie supernovae catalogue from the DES \citep{Popovic2025}, which comprises of approximately 1600 supernovae from 5 years of observations by the DES and a compilation of approximately 200 low-redshift supernovae from other surveys. The supernovae measurements give us a measure of the distance modulus
\begin{align} 
    \mu = 5\log_{10}D_{\mathrm{L}}+25, 
\end{align}
where $D_{\mathrm{L}}$ is the luminosity distance given by 
\begin{equation}
    D_{\mathrm{L}}(z)=(1+z)D_{\mathrm{M}}(z).
\end{equation}
For observational predictions used in the likelihood we compute the theoretical luminosity distance for each supernovae as
\begin{equation}
    D_{\mathrm{L}}(z_{\mathrm{Hel}},z_{\mathrm{CMB}})=(1+z_{\mathrm{Hel}})(1+z_{\mathrm{CMB}})D_{\mathrm{A}}(z_{\mathrm{CMB}})
\end{equation}
given in units of $\mathrm{Mpc}$, where $D_{\mathrm{A}}$ is the angular diameter distance
\begin{equation}
    D_{\mathrm{A}}(z)=\frac{D_{\mathrm{M}}(z)}{(1+z)}
\end{equation}
and $z_{\mathrm{Hel}}$ the Heliocentric redshift, measured in the rest frame of the Sun, of the supernovae and $z_{\mathrm{CMB}}$ the redshift of the supernovae transferred into the CMB rest frame. The likelihood follows the same form as the Gaussian likelihood shown in Eq.~\ref{eq_cmb_likelihood}. We use the full covariance matrix provided in the DES-Dovekie dataset, which incorporates both statistical and systematic uncertainties and their correlations across the supernova sample. We will denote constraints from the DES-Dovekie supernovae with the shorthand SN.

\subsubsection{Positive ISW}

Measurements of the ISW effect are inherently challenging because it contributes primarily on large angular scales, where cosmic variance is large, and becomes rapidly overwhelmed by the primordial CMB anisotropies at $\ell \gtrsim 50$. However, \citet{Renk2017} and \citet{Seraille2024} have shown that weak ISW measurements are still a powerful probe of modified gravity, due to the sensitivity of the $1-f_{1}-\zeta$ term within the ISW integral. Typically, $1-f_{1}(z)>0$, but the appearance of the extra $\zeta$ term can cause this term to flip sign. This change of sign can only be measured in large-scale structure to ISW cross-correlation measurements made by cross-correlating galaxy surveys with the CMB. The power of this measurement can be seen in \citet{Renk2017}, where galaxy-ISW cross-correlation measurements have been used to rule out CG models with $f_{\phi}=1$. Furthermore, \citet{Seraille2024} show that the ISW greatly constrains the amplitude of any modifications to the lensing potential $\Sigma$. One can see this emerges directly from the definition of $\zeta$ (see Eq.~\ref{eq_zeta}), which is sensitive to the temporal gradient of $\Sigma$. Almost all galaxy-ISW cross-correlation measurements made to date have measured positive values, meaning any model that predicts a large time gradient in the modification to the lensing potential will be ruled out by these observations.

To constrain models with the ISW using galaxy-ISW cross-correlation measurements would typically require predicting the matter power spectrum and including a galaxy bias prescription. Since we leave constraints that require predictions of the matter power spectrum for a future analysis, we instead apply the strong discriminatory power of the ISW by imposing a prior on the ISW strength $S_{\mathrm{ISW}}$ (see Eq.~\ref{eq_ISW_strength}) that ensures the ISW remains positive. The likelihood to maintain positive ISW cross-correlations is written as a prior

To fully constrain these models using galaxy-ISW cross-correlation measurements would require predictions for the matter power spectrum together with a galaxy bias prescription. Since we leave such an analysis to future work, we instead impose the weaker requirement that the predicted galaxy-ISW cross-correlation remains positive, consistent with current measurements. This should therefore be interpreted as a sign-consistency prior rather than a full statistical ISW likelihood, and does not make use of the measured amplitudes or uncertainties of existing galaxy-ISW cross-correlations.
\begin{equation}
    \log\mathcal{L}_{\mathrm{ISW}}=\begin{dcases}
        \,0, \quad &\text{if }S_{\mathrm{ISW}}\ge 0,\\
        \,-\infty, \quad &\text{otherwise,}
    \end{dcases}
\end{equation}
with the associated constraints denoted with the shorthand ISW.

\subsection{Inference of Model Parameters}

We infer posterior probability distribution for the model parameters in a Bayesian framework. The posterior probability distribution is given by Bayes' theorem
\begin{equation}
    P(\theta|D) = \frac{\mathcal{L}(D|\theta)\,\pi(\theta)}{Z},
\end{equation}
where $\theta$ denotes the model parameters and $D$ the data. The likelihood is expressed as $\mathcal{L}(D| \theta)$, the prior $\pi(\theta)$, and the Bayesian evidence $Z$, which acts as a normalization constant relevant for model comparison.

We constrain the parameters of the two models presented in Eqs.~\ref{Kdef} and \ref{Gdef}, `Growing $\mathcal{G}(\phi)$' and `Decaying $\mathcal{K}(\phi)$'. We compare the constraints of these models to constraints on $\Lambda$CDM and dynamical DE ($w_{0}w_{a}$CDM). For all the models we vary the physical densities of baryons $w_{\mathrm{b}}=\Omega_{\mathrm{b},0}h^{2}$ and CDM $w_{\mathrm{c}}=\Omega_{\mathrm{c},0}h^{2}$. Our initial conditions also require us to vary the Hubble constant of a reference cosmology from which our models depart at an early redshift (typically we set this to be at $z_{\rm ini}\approx1000$), see Appendix \ref{ICs} for more details . This reference $H_0$ value is denoted $H_{0}^{\mathrm{GR}}$. For $\Lambda$CDM and $w_{0}w_{a}$CDM, $H_{0}^{\mathrm{GR}}$ plays its normal role.  For the ACG models, we additionally sample the ratio between the contribution from the scalar field over $\Lambda$, given by the term $f_{\phi}^{\mathrm{in}}$ (defined in Eq.~\ref{eq_f_phi}) at $z_{\rm ini}$, and the constant $c_{g_{3}}$ for the Growing $\mathcal{G}(\phi)$ and $c_{k}$ for the Decaying $\mathcal{K}(\phi)$. 

In our main analysis, we fix the sum of the masses of neutrino species to $\sum m_{\nu}=0.06\,\mathrm{eV}$. However, as a consistency test, we allow the effective mass of neutrinos $\sum m_{\nu}^{\mathrm{eff}}$ to be an additional free parameter. Here, we substitute the mass of neutrinos with an effective mass that is allowed to be negative (see Sec.~\ref{sec_mass_neu}). This allows us to test whether the ACG models are able to alleviate the preference for negative neutrino mass that is present in constraints from $\Lambda$CDM that is reduced in $w_{0}w_{a}$CDM.

The conventional shift symmetric CG and ESS models (see Table~\ref{tab_horndeski_models}) produce an EoS for DE that is always phantom, making them difficult to reconcile with current observations. For this reason, we find constraints are often pulled towards $\Lambda$CDM and an $f_{\phi}\rightarrow 0$. We discuss these models and their parameter constraints in more detail in Appendix~\ref{app_shift_symmetric}.

\begin{table}
    \centering
    \begin{tabular}{llll}
        \hline\hline
        Model & Parameter & Prior Range & Fixed Value\\
        \hline\hline
        \multirow{3}{*}{All models} & $H_{0}^{\mathrm{GR}}$ & $[50, 100]$ & --\\
        & $\Omega_{\mathrm{b},0}h^{2}$ & $[0.02, 0.025]$  & --\\
        & $\Omega_{\mathrm{c},0}h^{2}$ & $[0.1, 0.15]$  & --\\
        \hline
        \multirow{2}{*}{Dynamical DE} & $w_{0}$ & $[-1.5, 0.5]$  & -1 \\
        & $w_{a}$ & $[-3, 0.5]$ & 0 \\
        \hline
        Growing $\mathcal{G}(\phi)$ \& & \multirow{2}{*}{$f_{\phi}^{\mathrm{in}}$} & \multirow{2}{*}{$[0,1]$} & \multirow{2}{*}{0}\\
        Decaying $\mathcal{K}(\phi)$ & & & \\
        \hline
        Growing $\mathcal{G}(\phi)$ only & $c_{g_{3}}$ & $[0, 50]$ & 0\\
        Decaying $\mathcal{K}(\phi)$ only & $c_{k}$ & $[0, 20]$ & 0\\
        \hline
        Extension to all models &  $\sum m_{\nu}^{\mathrm{eff}}$ & $[-1,1]$ & 0.06\\
        \hline\hline
    \end{tabular}
    \caption{We list the free model parameters used in our analysis, their prior ranges (given as [minimum, maximum]) and their corresponding fixed values. The first three parameters are varied in all models. The DE parameters $w_{0}$ and $w_{a}$ are varied only in analyses with dynamical DE. The parameter $f_{\phi}^{\mathrm{in}}$ is free in both ACG models, while $c_{g_{3}}$ and $c_{k}$ are varied exclusively in the Growing $\mathcal{G}(\phi)$ and Decaying $\mathcal{K}(\phi)$ models, respectively. The effective neutrino mass, $\sum m_{\nu}^{\mathrm{eff}}$, is initially fixed and subsequently allowed to vary as a consistency check in all models.}
    \label{tab_parameter_priors}
\end{table}

\begin{figure*}
    \centering
    \includegraphics[width=0.8\textwidth]{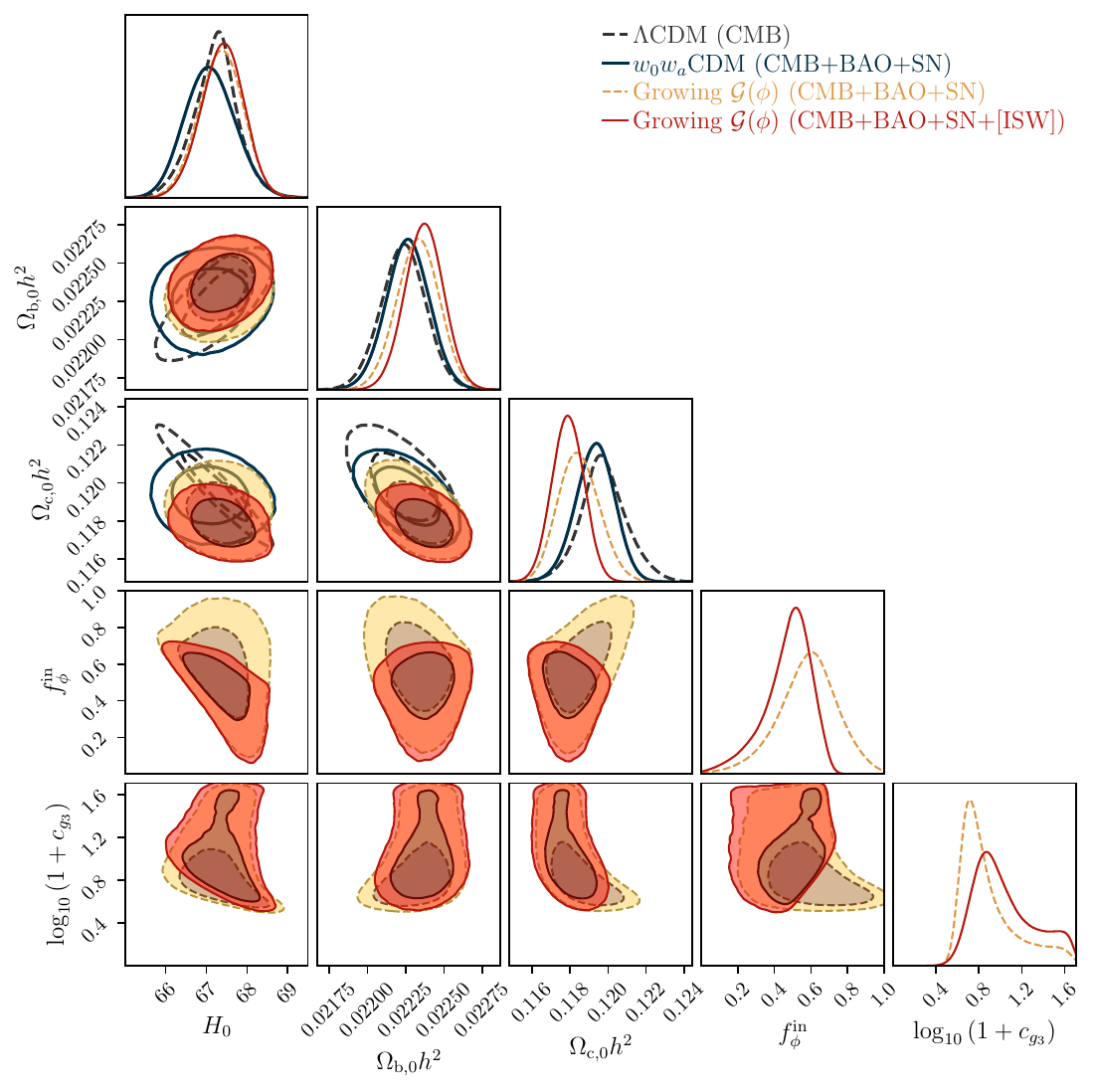}
    \caption{Posterior constraints on the Growing $\mathcal{G}(\phi)$ model with CMB, BAO and SN observations with (red contours) and without (yellow contours) a positive ISW prior shown in comparison to constraints from $\Lambda$CDM (from CMB-only; black dashed lines) and $w_{0}w_{a}$CDM (dark blue solid lines). We show the constraints on Hubble expansion rate $H_{0}$, baryon density $\Omega_{\mathrm{b},0}h^{2}$ and CDM density $\Omega_{\mathrm{c},0}h^{2}$ for all the models, showing generally consistent constraints on $H_{0}$, while Growing $\mathcal{G}(\phi)$ shows a preference for larger baryon density and smaller CDM density.  The positive ISW prior eliminates large $f_{\phi}^{\mathrm{in}}$ in favour of moderate $f_{\phi}^{\mathrm{in}}$ and larger and broader constraints on $c_{g_{3}}$.}
    \label{fig_G3_constraints}
\end{figure*}

In Table~\ref{tab_parameter_priors}, we list the models we evaluate in this analysis, together with their associated free parameters; including their prior ranges and, where applicable, their fixed fiducial values.

We use the nested sampler \texttt{DYNESTY} \citep{Speagle2020} to compute the Bayesian evidence $Z$ for each model. The nested-sampling runs are deemed converged once the estimated uncertainty on the log-evidence reaches $\Delta \ln Z = 0.01$. To obtain robust estimates of the posterior distributions and parameter constraints, we additionally sample the posterior using the affine-invariant ensemble sampler \texttt{emcee} \citep{Foreman2013}. Unless otherwise stated, all posterior distributions and confidence contours shown in this work, are derived from the \texttt{emcee} chains, while Bayesian evidence values are obtained from \texttt{DYNESTY}.

\section{Constraints and Model Comparison}
\label{results}

In this section, we present constraints on the ACG models in comparison to $\Lambda$CDM and $w_{0}w_{a}$CDM. In particular, we discuss constraints on model parameters, relative fits to data, implications for the effective EoS and, lastly, implications for the growth of structure. 

\begin{figure*}
    \centering
    \includegraphics[width=0.8\textwidth]{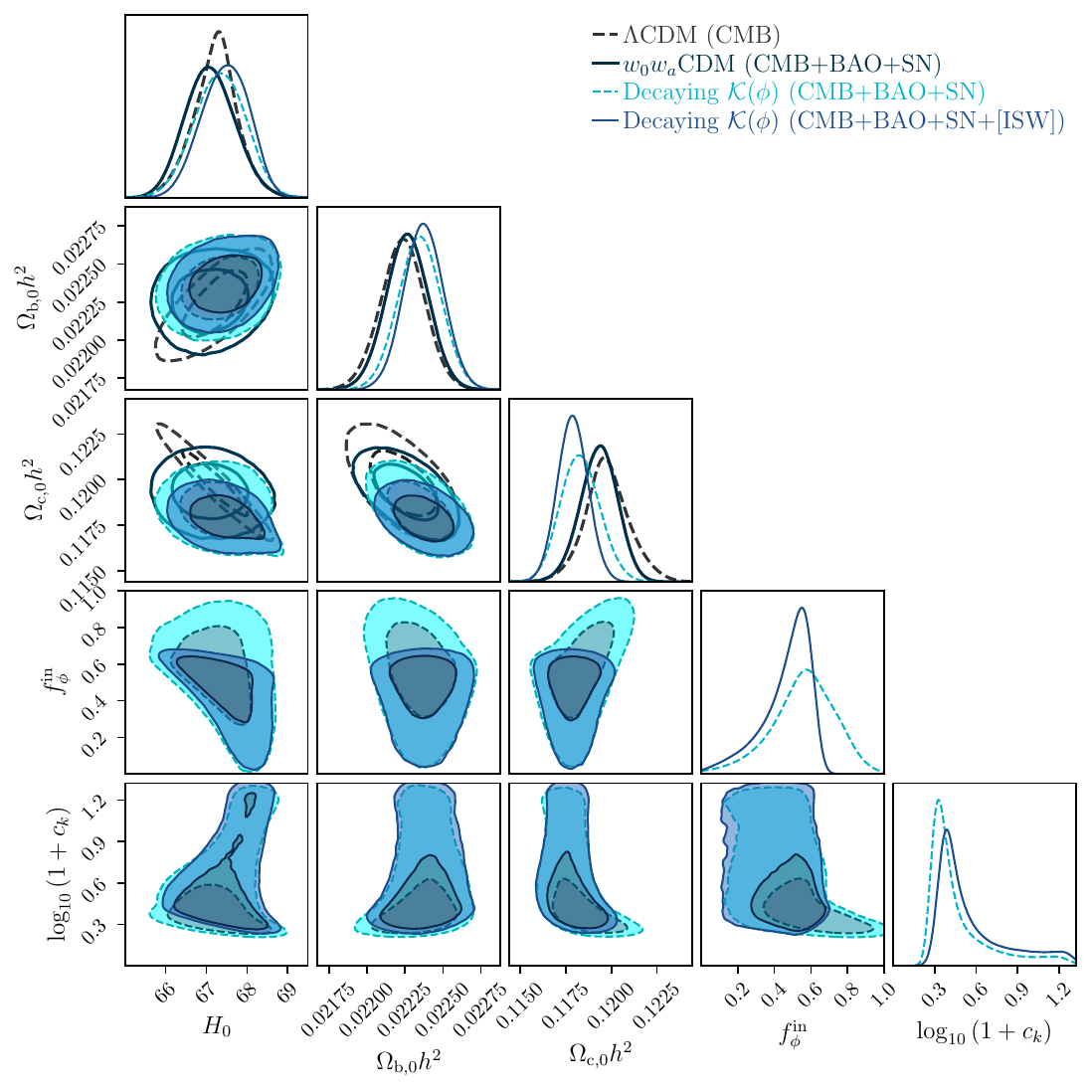}
    \caption{Posterior constraints on the Decaying $\mathcal{K}(\phi)$ model with CMB, BAO and SN observations with (dark blue contours) and without (light blue contours) a positive ISW prior in comparison to constraints from $\Lambda$CDM (from CMB-only; black dashed lines) and $w_{0}w_{a}$CDM (dark blue solid lines). We show the constraints on Hubble expansion rate $H_{0}$, baryon density $\Omega_{\mathrm{b},0}h^{2}$ and CDM density $\Omega_{\mathrm{c},0}h^{2}$ for all the models, showing generally consistent constraints on $H_{0}$, while Growing $\mathcal{K}(\phi)$ shows a preference for larger baryon density and smaller CDM density. The positive ISW prior eliminates large $f_{\phi}^{\mathrm{in}}$ in favour of moderate $f_{\phi}^{\mathrm{in}}$ and larger and broader constraints on $c_{k}$.}
    \label{fig_K_constraints}
\end{figure*}

\subsection{Parameter Constraints}
\label{sec_sub_constraints}

We reproduce parameter constraints on $\Lambda$CDM and dynamical DE using combinations of the observations listed in section \ref{observations}, and compare them to our Growing $\mathcal{G}(\phi)$ and Decaying $\mathcal{K}(\phi)$ ACG models.

In Fig.~\ref{fig_G3_constraints}, we compare $\Lambda$CDM constraints from the CMB-only (black dashed line) -- since CMB and DESI BAO DR2 are in tension in $\Lambda$CDM \citep[see][]{DESI_BAO_DR2,Hergt2026} -- to joint constraints from the CMB, BAO and SN for dynamical DE (dark blue solid line), and the Growing $\mathcal{G}(\phi)$ ACG model with (red contours) and without (yellow contours) a positive ISW prior. The Growing $\mathcal{G}(\phi)$ model shows fairly consistent constraints on $H_{0}$, but with a mild preference for larger baryon density $\Omega_{\mathrm{b},0}h^{2}$ and a lower CDM density; that is exaggerated by adding the positive ISW prior. However, the shift in the baryon and CDM densities appears to be limited to the baryon-CDM ratio, since the overall matter density today $\Omega_{\mathrm{m},0}$ for the Growing $\mathcal{G}(\phi)$ is consistent with dynamical DE but larger than $\Lambda$CDM (see Table~\ref{tab_constraints}). 

The Growing $\mathcal{G}(\phi)$ parameters $f_{\phi}^\mathrm{in}$ and $c_{g_{3}}$ show a clear preference away from the $\Lambda$CDM limits ($f_{\phi}^\mathrm{in}\!=\!0$ and $c_{g_{3}}\!=\!0$), with $f_{\phi}^\mathrm{in}$ peaking around $\sim0.65$ and $c_{g_{3}}$ peaking around $\sim4$. The large tail in the posterior of $c_{g_{3}}$ is due to a diminishing sensitivity to larger values in the dynamics of the Growing $\mathcal{G}(\phi)$ model, and for this reason we plot 
$\log_{10}(1+c_{g_{3}})$. The impact of the positive ISW prior is best illustrated in the subpanel showcasing the posteriors of $f_{\phi}^{\mathrm{in}}$ in relation to $c_{g_{3}}$, where we see the ISW prior cuts the posterior, eliminating regions of the parameter space with high $f_{\phi}^{\mathrm{in}}$ in favour of moderate $f_{\phi}^{\mathrm{in}}\approx0.5$ and larger and broader constraints on $c_{g_{3}}$. In Fig.~\ref{fig_K_constraints}, we show the parameter constraints on the Decaying $\mathcal{K}(\phi)$ model, that follows a similar relation to the constraints to the Growing $\mathcal{G}(\phi)$ model. This behaviour is expected due to a degeneracy between the models \citep{Hallam2026}.
\begin{figure*}
    \centering
    \includegraphics[width=0.8\textwidth]{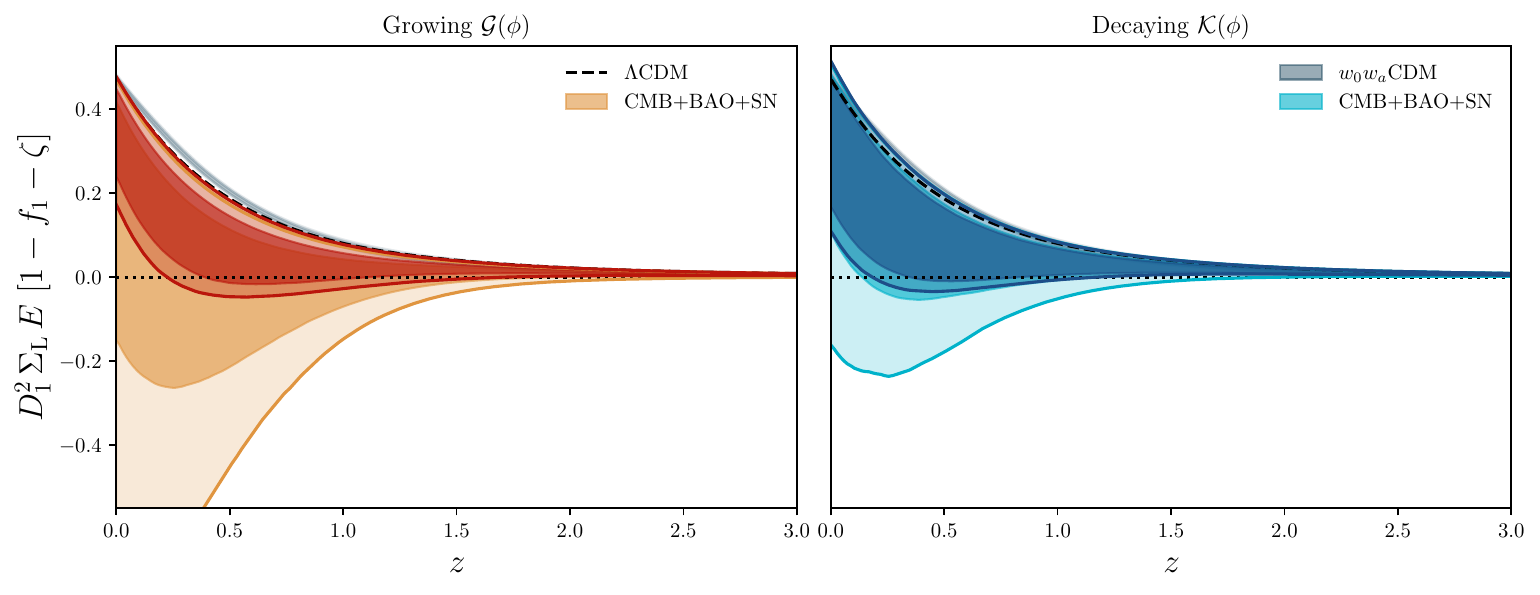}
    \caption{Constraints on the ISW strength integral (Eq.~\ref{eq_ISW_strength}) for the Growing $\mathcal{G}(\phi)$ model (on the left) and the Decaying $\mathcal{K}(\phi)$ model (on the right) in comparison to $\Lambda$CDM (dashed black line) and $w_{0}w_{a}$CDM (grey). The ACG models are shown with (red and dark blue) and without (yellow and pale blue) the positive ISW prior for the Growing $\mathcal{G}(\phi)$ model and the Decaying $\mathcal{K}(\phi)$ model respectively. Although the positive ISW prior removes regions of the parameter space giving negative ISW cross-correlations, both ACG models predict smaller ISW cross-correlations than is expected from $\Lambda$CDM or $w_{0}w_{a}$CDM.}
    \label{fig_ISW_integral}
\end{figure*}

In Fig.~\ref{fig_ISW_integral}, we show the impact of the constraints with and without the positive ISW prior on the integral of the ISW strength (see Eq.~\ref{eq_ISW_strength}), showing that the Growing $\mathcal{G}(\phi)$ model has a broader range, allowing for the integral function to be more deeply negative. The constraints on the integral function shows these models predict an ISW integral function that is below the $\Lambda$CDM and $w_{0}w_{a}$CDM expectation. If these models are capturing the true behaviour of DE, much smaller ISW-galaxy cross-correlation amplitudes are expected to be observed as compared to $\Lambda$CDM.

\begin{table*}
    \centering
    \begin{tabular}{lllllll}
        \hline\hline
        \T\B Parameters & $\Lambda\mathrm{CDM}$ & $w_{0}w_{a}\mathrm{CDM}$ & $\textrm{Growing}\;\mathcal{G}(\phi)$ & $...+\textrm{[ISW]}$ & $\textrm{Decaying}\;\mathcal{K}(\phi)$ & $...+\textrm{[ISW]}$ \\
        \hline\hline
        \T\B $H_{0}^{\mathrm{GR}}$ & $68.14\pm0.32$ & $67.17\pm0.58$ & $82.2^{+4.0}_{-5.1}$ & $78.4^{+6.3}_{-4.0}$ & $82.1^{+5.7}_{-5.2}$ & $82.0^{+5.2}_{-3.5}$ \\
        \B $10^{2}\,\Omega_{\mathrm{b},0}h^{2}$ & $2.238\pm0.012$ & $2.227\pm0.014$ & $2.233 \pm 0.012$ & $2.237\pm 0.013$ & $2.24\pm0.012$ & $2.24\pm0.014$ \\
        \B $10^{2}\,\Omega_{\mathrm{c},0}h^{2}$ & $11.77\pm0.07$ & $11.93\pm0.10$ & $11.85\pm 0.10$ & $11.78\pm 0.08$ & $11.82\pm 0.09$ & $11.796\pm0.079$\\
        \B $w_{0}$ & -- & $-0.79\pm 0.06$ & -- & -- & -- & -- \\
        \B $w_{a}$ & -- & $-0.71^{+0.19}_{-0.32}$ & -- & -- & -- & -- \\
        \B $f_{\phi}^{\mathrm{in}}$ & -- & -- & $0.67^{+0.12}_{-0.20}$ & $0.49\pm 0.11$ & $0.56^{+0.12}_{-0.17}$ & $0.581^{+0.068}_{-0.118}$\\
        \B $c_{g_{3}}$ & -- & -- & $4.4^{+3.7}_{-2.0}$ & $6.9^{+12.7}_{-5.1}$ & -- & -- \\
        \B $c_{k}$ & -- & -- & -- & -- & $1.4^{+1.6}_{-1.1}$ & $1.55^{+1.03}_{-0.79}$ \\
        \hline
        \T\B $H_{0}$ & $68.14\pm 0.32$ & $67.17 \pm 0.58$ & $67.29\pm0.53$ & $67.36^{+0.51}_{-0.44}$ & $67.30^{+0.69}_{-0.48}$ & $67.29^{+0.57}_{-0.62}$ \\
        \B $10^{2}\,\Omega_{\mathrm{m},0}$ & $30.32\pm0.42$ & $31.53 \pm 0.59$ & $31.24\pm0.59$ & $31\pm 0.52$ & $31.07^{+5.0}_{-6.9}$ & $31.25^{+5.0}_{-7.2}$ \\
        \B $f_{\phi,0}$ & -- & -- & $0.103^{+0.359}_{-0.082}$ & $0.113^{+0.097}_{-0.090}$ & $0.16^{+0.20}_{-0.14}$ & $0.17^{+0.095}_{-0.144}$ \\
        \hline
        \hline
    \end{tabular}
    \caption{Parameter constraints on $\Lambda$CDM, $w_{0}w_{a}$CDM, Growing $\mathcal{G}(\phi)$ and Decaying $\mathcal{K}(\phi)$ from observations of the CMB, BAO and SN. For ACG models, we additionally provide constraints with a positive ISW prior (denoted with a +[ISW]). The free input parameters are listed in the first top section of the table, with derived parameters shown in the bottom section. $H_{0}^{\mathrm{GR}}$, $\Omega_{\mathrm{b},0}h^{2}$ and $\Omega_{\mathrm{c},0}h^{2}$ are free parameters in all the models, while $w_{0}$ and $w_{a}$ are only free in dynamical DE, $f_{\phi}^{\mathrm{in}}$ only for ACG models, and $c_{g_{3}}$ and $c_{k}$ for Growing $\mathcal{G}(\phi)$ and Decaying $\mathcal{K}(\phi)$ models respectively. In addition, we provide constraints on the derived Hubble expansion rate $H_{0}$, matter density $\Omega_{\mathrm{m},0}$ and the fraction of the scalar field to DE ratio today $f_{\phi,0}$. For $\Lambda$CDM and  $w_{0}w_{a}$CDM, $H_0=H_{0}^{\mathrm{GR}}$.}
    \label{tab_constraints}
\end{table*}

In Table~\ref{tab_constraints}, we summarise constraints from observations of the CMB, BAO and SN on parameters from $\Lambda$CDM, $w_{0}w_{a}$CDM, Growing $\mathcal{G}(\phi)$ and Decaying $\mathcal{K}(\phi)$. We show constraints on the input free parameters $H_{0}^{\mathrm{GR}}$, $\Omega_{\mathrm{b},0}h^{2}$ and $\Omega_{\mathrm{c},0}h^{2}$ for all models, $w_{0}$ and $w_{a}$ for dynamical DE, $f_{\phi}^{\mathrm{in}}$ for the ACG models, and $c_{g_{3}}$ and $c_{k}$ for the Growing $\mathcal{G}(\phi)$ and Decaying $\mathcal{K}(\phi)$ models respectively. In addition, we provide constraints on the derived Hubble expansion rate $H_{0}$, matter density $\Omega_{\mathrm{m},0}$ and the fraction of the scalar field to DE ratio today $f_{\phi,0}$.

\subsection{Model Comparison}
\label{sec_sub_comparison}
\begin{table*}
    \centering
    \begin{tabular}{lcccccc}
        \hline\hline
        \T\B & $\Lambda\mathrm{CDM}$ & $w_{0}w_{a}\mathrm{CDM}$ & $\textrm{Growing}\;\mathcal{G}(\phi)$ & $...+\textrm{[ISW]}$ & $\textrm{Decaying}\;\mathcal{K}(\phi)$ & $...+\textrm{[ISW]}$ \\
        \hline\hline
        \T\B $\Delta \chi_{\mathrm{MAP}}^{2}$ & 0 & $-13.3$ & $-11.4$ & $-10.5$ & $-11.2$ & $-10.6$ \\
        \B Significance & 0 & $3.01\sigma$ & $2.71\sigma$ & $2.57\sigma$ & $2.68\sigma$ & $2.58\sigma$ \\
        \hline
        \T $\Delta \ln Z$ & 0 & $1.72$ & $2.95$ & $2.09$ & $1.52$ & $2.27$ \\
        \B Betting Odds & 1 & $5.67$ & $19.1$ & $8.09$ & $4.59$ & $9.68$ \\
        \hline
        \hline
    \end{tabular}
    \caption{We compare the improvement in $\chi^{2}$ and Bayesian evidence $Z$ for dynamical DE and the Growing $\mathcal{G}(\phi)$ and Decaying $\mathcal{K}(\phi)$ ACG models with respect to $\Lambda$CDM. For $\chi^{2}$, we compare the MAP values $\Delta\chi^{2}_{\mathrm{MAP}}$ and convert the shift into a significance assuming a $\chi^{2}$ distribution with 2 degrees of freedom for the 2 extra free parameters in comparison to $\Lambda$CDM. The Bayesian evidence is converted into betting odds, which shows the relative preference for a given model with respect to $\Lambda$CDM. Dynamical DE shows the most significance shift in the $\chi^{2}_{\mathrm{MAP}}$ with the ACG models providing a more significant improvement in the Bayesian evidence (specifically when including the positive ISW prior), showing that although dynamical DE provides the best fit, the model is penalised for being less predictive than the ACG models considered.}
    \label{tab_model_comparison}
\end{table*}

\begin{figure*}
    \centering
    \includegraphics[width=0.8\textwidth]{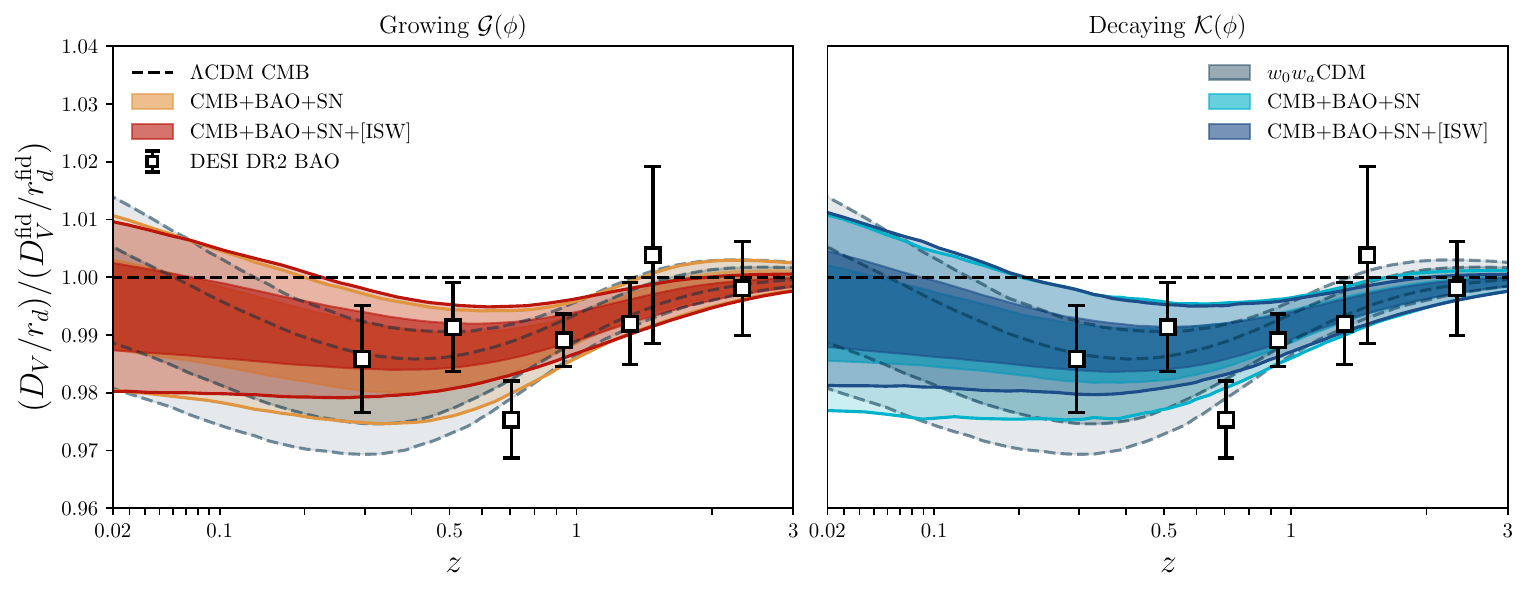}
    \caption{Volume average BAO measurements $D_{\mathrm{V}}/r_{\mathrm{d}}$ from DESI DR2 against constraints on dynamical DE and the ACG models in comparison to the fiducial Planck $\Lambda$CDM model. Dynamical DE (grey) and the ACG Growing $\mathcal{G}(\phi)$ (left; red) and Decaying $\mathcal{K}(\phi)$ (right; blue) models are shown from the joint constraints from the Planck CMB, DESI BAO and DES-Dovekie supernovae. The dark shades (red and dark blue) indicate the ACG models constraints with the additional positive ISW prior. Dynamical DE and ACG favour a dip in the volume average BAO over $\Lambda$CDM, with the ACG models (in particular with the positive ISW prior), preferring shallower profiles.}
    \label{fig_BAO}
\end{figure*}

\begin{figure*}
    \centering
    \includegraphics[width=0.8\textwidth]{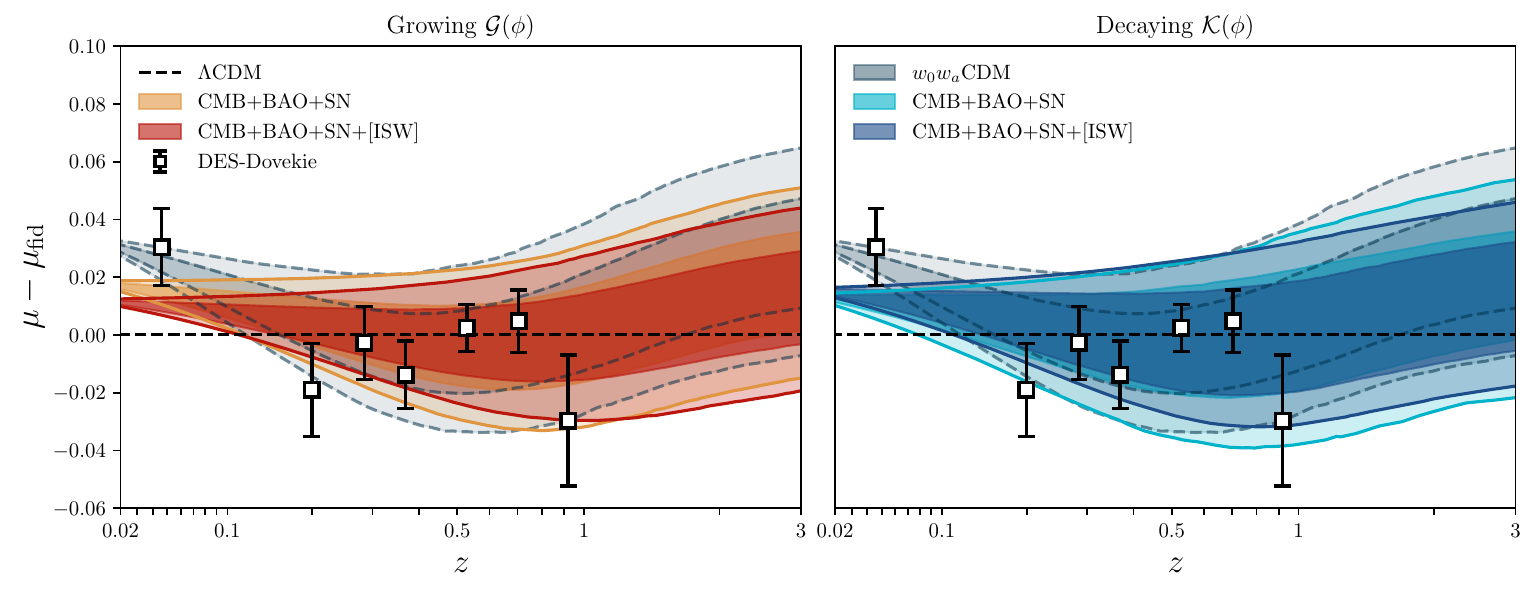}
    \caption{Distance modulus $\mu$ from DES-Dovekie measurements against constraints on dynamical DE and the ACG models in comparison to the fiducial Planck $\Lambda$CDM model. This figure uses the same colour scheme as Fig.~\ref{fig_BAO}. ACG models favour smaller values in the distance modules than dynamical DE at low redshift, albeit larger than what is predicted from $\Lambda$CDM.}
    \label{fig_distance_modulus}
\end{figure*}

In \citet{DESI_BAO_DR2} it was shown that observations from the CMB, DESI BAO and SN (particularly from DES) favour dynamical DE over $\Lambda$CDM. This was characterised by measuring the difference between the maximum a posteriori (MAP) estimates of the $\chi^{2}$, referred to as $\chi^{2}_{\mathrm{MAP}}$, which is computed by taking
\begin{equation}
    \chi^{2}_{\mathrm{MAP}}=(\vec{d}_{\mathrm{MAP}}-\vec{d}_{\mathrm{obs}})^{T}\cdot \mat{C}_{\mathrm{obs}}\cdot (\vec{d}_{\mathrm{MAP}}-\vec{d}_{\mathrm{obs}}),
\end{equation}
where subscript $\mathrm{MAP}$ denotes the MAP prediction for the observed data vector $\vec{d}_{\mathrm{obs}}$ and covariance $\mat{C}_{\mathrm{obs}}$. For combined constraints we simply add $\chi^{2}$ from the different observations. We compare the $\chi^{2}_{\mathrm{MAP}}$ of $w_{0}w_{a}$CDM and ACG models with respect to $\Lambda$CDM
\begin{equation}
    \Delta \chi^{2}_{\mathrm{MAP}} = \chi^{2}_{\mathrm{MAP}}-\chi^{2}_{\mathrm{MAP}-\Lambda\mathrm{CDM}}.
\end{equation}
The $\Delta \chi^{2}_{\mathrm{MAP}}$ can be converted into a significance assuming the difference follows a $\chi^{2}$ distribution with two degrees of freedom, this is because the dynamical DE and the ACG models all require the addition of two free parameters. In addition to $\Delta \chi^{2}_{\mathrm{MAP}}$ comparisons, \citet{DESI_BAO_DR2} also provide measures of the Deviance Information Criterion (DIC). However, this model comparison metric only holds when parameter constraints are Gaussian; for the ACG models the constraints on $f_{\phi}^{\mathrm{in}}$ and, in particular, the constants $c_{g_{3}}$ and $c_{k}$ are very non-Gaussian and the DIC fails to produce a reliable metric. 

The $\chi^{2}_{\mathrm{MAP}}$ and its associate significance is a frequentist model comparison metric; for a Bayesian approach we compare the Bayesian evidence $Z$. Bayesian evidence measures how well the model explains the data, penalising over complex models by measuring whether a model spreads the posterior mass over a large parameter space. We compute the Bayesian evidence using the nested sampler \texttt{DYNESTY} and compute the Bayesian evidence ratio
\begin{equation}
    \Delta\ln Z=\ln Z-\ln Z_{\Lambda\mathrm{CDM}}
\end{equation}
where $\ln Z_{\Lambda\mathrm{CDM}}$ is the Bayesian evidence measured for $\Lambda$CDM. The Bayesian evidence can be turned into betting odds by computing $\exp(\Delta\ln Z)$, which gives us a measure of the betting odds in favour of a given model with respect to $\Lambda$CDM.

In Table~\ref{tab_model_comparison}, we compare the $\Delta \chi^{2}_{\mathrm{MAP}}$ and Bayesian evidence for dynamical DE and the ACG models to $\Lambda$CDM. We show that the $\Delta \chi^{2}_{\mathrm{MAP}}$ favours $w_{0}w_{a}$CDM over the ACG models, with $w_{0}w_{a}$CDM showing $\Delta \chi^{2}_{\mathrm{MAP}}\sim -13$ in comparison to $\Delta \chi^{2}_{\mathrm{MAP}}\sim -11$ for the ACG models, but with both models clearly favoured over $\Lambda$CDM. For the Bayesian evidence, we find a modest increase in the preference for the ACG models, particularly when the positive ISW prior is imposed, yielding $\Delta \ln Z \simeq 2$ compared to $\Delta \ln Z \simeq 1.7$ for $w_{0}w_{a}$CDM. Although both models introduce the same number of additional parameters relative to $\Lambda$CDM, the ACG models are characterised by a more restrictive and less flexible EoS evolution. In particular, the Horndeski dynamics constrain the allowed behaviour of $w_{\mathrm{DE}}$, whereas the $w_{0}w_{a}$CDM parametrisation permits a broader range of phenomenological EoS trajectories. Consequently, a larger fraction of the $w_{0}w_{a}$CDM parameter volume may correspond to regions that are weakly supported by the data, leading to a stronger Occam penalty in the Bayesian evidence.

\subsection{Observables and the effective Equation of State}
\label{sec_sub_EoS}

To understand how the different models fit observations of DESI DR2 BAO and DES-Dovekie supernovae, we compare the observational measurements to predictions from the posterior constraint. In Fig.~\ref{fig_BAO}, we plot the volume average BAO from DESI with respect to the mean $\Lambda$CDM prediction from Planck-only, in comparison to dynamical DE and the ACG models favoured by joint constraints from Planck, DESI BAO and DES-Dovekie supernovae. The volume average BAO shows a clear suppression at low-redshift with respect to $\Lambda$CDM. The dynamical DE fits show the most significant dip, with the ACG models showing a more gradual depression, an effect which is more exaggerated once the positive ISW prior is added. In Fig.~\ref{fig_distance_modulus}, we plot the distance modulus with respect to $\Lambda$CDM for the same models (following \citealt{DESI_BAO_DR2} in marginalising over $H_{0}$). The main difference between dynamical DE is in the fit to the lowest redshift point where the ACG models show a preference for lower deviations from $\Lambda$CDM than dynamical DE. 

\begin{figure*}
    \centering
    \includegraphics[width=0.8\textwidth]{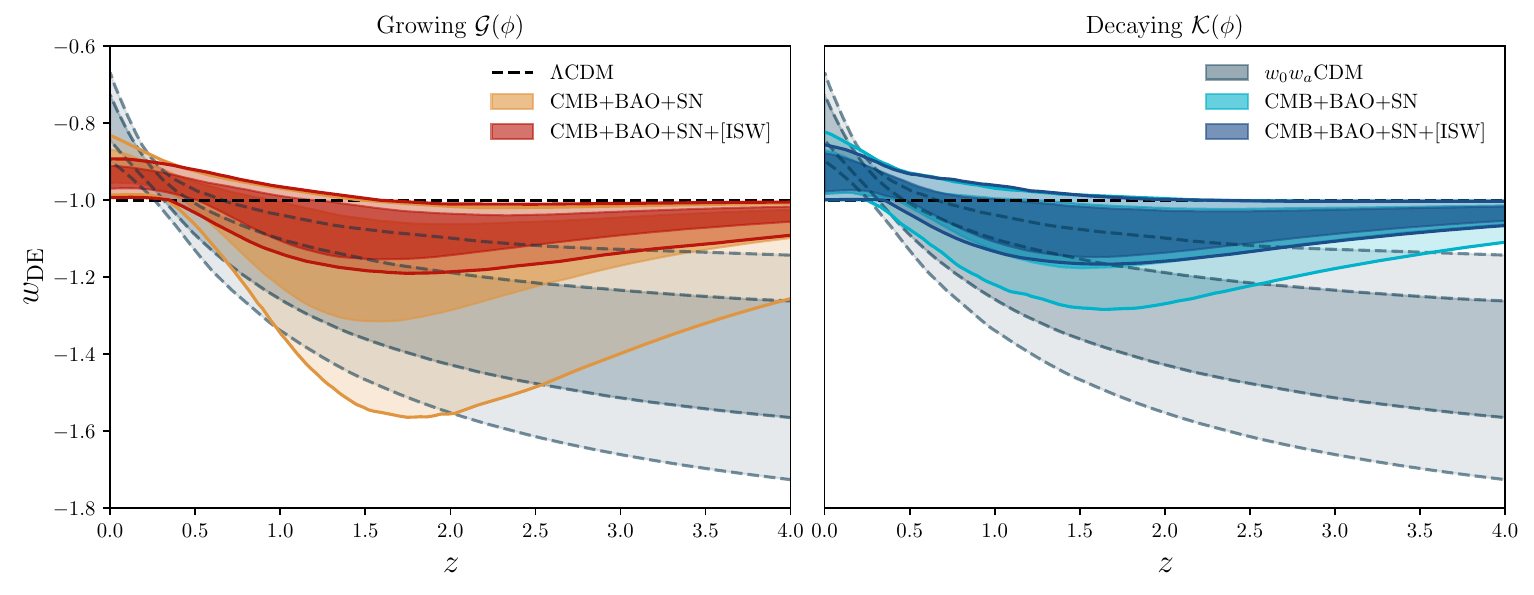}
    \caption{EoS for DE shown for dynamical DE and ACG models from the joint constraints of the CMB, DESI BAO and DES-Dovekie Supernovae. This figure uses the same colour scheme as Fig.~\ref{fig_BAO}. In contrast to dynamical DE, the EoS for both ACG models tends towards $-1$ at high redshift, and moves to a minima at around $z\sim 1.5$ before growing and crossing the phantom divide at redshift $z\sim0.7$. The Growing $\mathcal{G}(\phi)$ allows for a lower minima than the Decaying $\mathcal{K}(\phi)$, but the models exhibit very similar profiles once we add the positive ISW prior, which enforces a shallower EoS for ACG models with a slighter earlier phantom crossing $z\sim1$ in comparison to dynamical DE's $z\sim0.5$.}
    \label{fig_EoS}
\end{figure*}

\begin{figure*}
    \centering
    \includegraphics[width=0.8\textwidth]{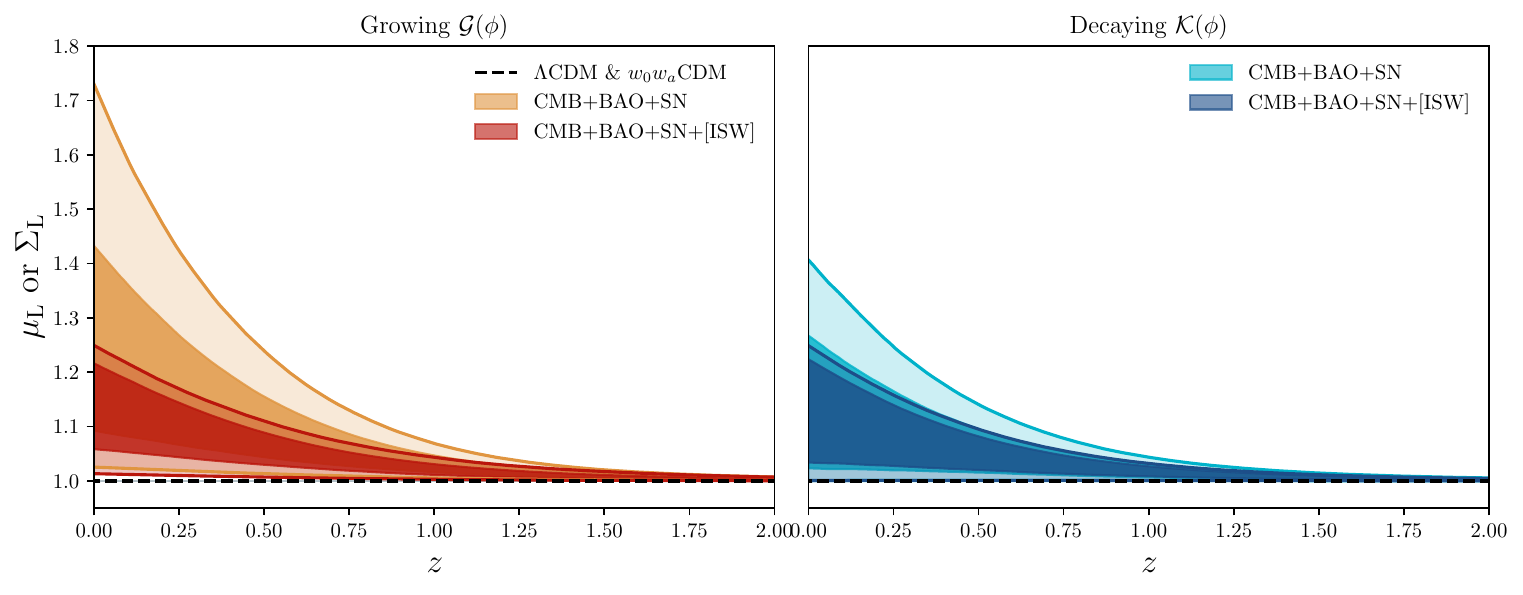}
    \caption{Linear modification to the Poisson equation $\mu_{\mathrm{L}}$ and lensing potential $\Sigma_{\mathrm{L}}$ for ACG models, where $\mu_{\mathrm{L}}=\Sigma_{\mathrm{L}}$. This figure uses the same colour scheme as Fig.~\ref{fig_BAO}. The positive ISW prior enforces shallower profiles on $\Sigma$, resulting in predictions for $\mu_{\mathrm{L}}$ and $\Sigma_{\mathrm{L}}$ that are much shallower, with steeper profiles for the Growing $\mathcal{G}(\phi)$ ruled out, resulting in broadly similar profiles for both models.}
    \label{fig_mu_sigma}
\end{figure*}

\begin{figure*}
    \centering
    \includegraphics[width=0.9\textwidth]{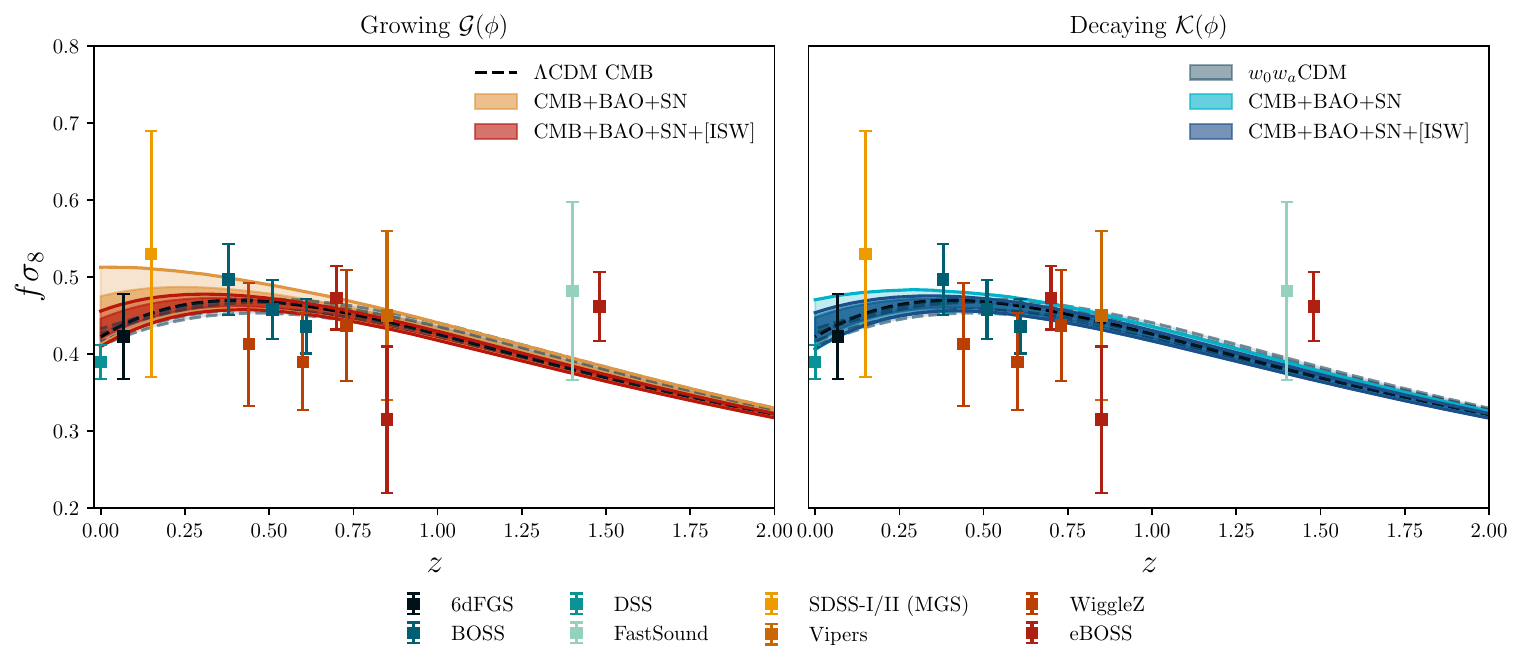}
    \caption{The growth of structure measured from a number of surveys in comparison to predictions from $\Lambda$CDM, dynamical DE and ACG models. This figure uses the same colour scheme as Fig.~\ref{fig_BAO}. We plot $f\sigma_{8}$, a product of the growth rate and time-evolution of $\sigma_{8}$. Dynamical DE and ACG models are shown for the joint constraints from Planck CMB, DESI BAO and DES-Dovekie SN. We compare this to the mean profiles from $\Lambda$CDM. Both the Growing $\mathcal{G}(\phi)$ and Decaying $\mathcal{K}(\phi)$ predict a moderate (but consistent with $\Lambda$CDM) increase in the growth of structure at low-redshift, with the ISW prior reducing the preference for slightly larger growth.}
    \label{fig_fsigma8}
\end{figure*}

In either case we see that the constraint from setting a prior on the ISW has only a small impact on the measured observables (see Fig.\ref{fig_BAO} and \ref{fig_distance_modulus}). However, while the change on the observational measurement is small, the change on the EoS is quite significant. In Fig.~\ref{fig_EoS}, we plot the EoS for the ACG models in comparison to the EoS of dynamical DE. The behaviour for the ACG models is very different to that observed from dynamical DE; firstly, the ACG models tend towards an EoS of -1 at high redshift, since the model is defined with a cosmological constant component to DE that is set with $f_{\phi}^{\mathrm{in}}$. Secondly, the ISW prior has a significant impact on the shape of the EoS, favouring an EoS that is much shallower than the EoS for dynamical DE. 

In Fig.~\ref{fig_mu_sigma}, we show the modification to the Poisson equation $\mu_{\mathrm{L}}$ and the modification to the lensing potential $\Sigma_{\mathrm{L}}$ defined in Eq.~\ref{eq_Sigma_L}, that are equal in the ACG model. Here, we see that the positive ISW prior restricts the allowed ranges in $\Sigma_{\mathrm{L}}$ to shallower profiles, since the sign of the ISW is dictated by the derivative of $\Sigma_{\mathrm{L}}$. A greater contribution from the scalar field to DE results in stronger $\Sigma_{\mathrm{L}}$ profiles. This means the ISW prior prefers models with a more significant contribution from the cosmological constant and therefore the scalar field has a shallower impact on quantities defining the combined DE contribution, such at the EoS of DE. The observables between the two models do not vary greatly, but the EoS for Growing $\mathcal{G}(\phi)$ can be much deeper than what is seen for Decaying $\mathcal{K}(\phi)$. However, the ISW prior limits the two models to an EoS that is shallower and that follows fairly identical profiles. Furthermore, one key difference between this model and dynamical DE is the moment of phantom crossing, that occurs generally at an earlier time than dynamical DE: $z\sim0.7-1$ in comparison to $z\sim0.5$.

\subsection{Growth of Structure}
\label{sec_sub_growth}

Although our constraints on ACG models are derived solely from the background expansion history, it is important to understand their impact on the growth of structure. Since ACG models closely recover $\Lambda$CDM at early times, they preserve the successful description of the primary CMB anisotropies and the corresponding initial conditions for structure formation. Consequently, any departures from $\Lambda$CDM are expected to arise primarily through the late-time evolution of cosmic structure. We therefore assume the same CMB-normalised initial amplitude of matter fluctuations as in $\Lambda$CDM,
\begin{equation}
    \sigma_{8}(z_{\rm early}) \approx \sigma_{8}^{\Lambda{\rm CDM}}(z_{\rm early}),
\end{equation}
where $z_{\rm early}$ denotes an epoch deep in the matter-dominated era, prior to the onset of DE dominance. In our analysis, we evaluate this condition at $z_{\mathrm{early}}=1200$. This coincides with the initial redshift of the solver, $z_{\mathrm{ini}}=1200$, although the two quantities are conceptually distinct: $z_{\mathrm{early}}$ denotes the epoch at which the $\sigma_{8}$ matching condition is imposed, while $z_{\mathrm{ini}}$ specifies the starting redshift for the numerical integration. The time evolution of $\sigma_{8}$ is given as a product of
\begin{equation}
    \sigma_{8}(z)=\sigma_{8}(z=0)\,D_{1}(z),
\end{equation}
where $D_{1}$ is the growth function. To fix $\sigma_{8}$ at early times we require
\begin{equation}
    \sigma_{8}(z=0)=0.8\times\,\frac{D_{1}^{\Lambda\mathrm{CDM}}(z_{\mathrm{early}})}{D_{1}(z_{\mathrm{early}})},
\end{equation}
where we assume $\sigma_{8}^{\Lambda\mathrm{CDM}}(z=0)=0.8$, approximately the value constrained by \citet{Planck2020}.  We take the mean value of $D_{1}^{\Lambda\mathrm{CDM}}(z_{\mathrm{early}})$ from constraints of $\Lambda$CDM from Planck-only measurements. In Fig.~\ref{fig_fsigma8}, we compare the predictive constraints on
\begin{equation}
    f\sigma_{8}=f_{1}(z)\,\sigma_{8}(z)
\end{equation}
for ACG models in comparison to $\Lambda$CDM and dynamical DE. We find the joint constraints from CMB, BAO and SN produce $f\sigma_{8}$ predictions that are broadly in agreement with the predictions of $\Lambda$CDM and dynamical DE, but show a preference for higher amplitudes at $z<0.5$. There is once again broad agreement between the Growing $\mathcal{G}(\phi)$ and Decaying $\mathcal{K}(\phi)$ models when the positive ISW prior is included. Without this prior, however, the Growing $\mathcal{G}(\phi)$ model permits larger low-redshift deviations. The results are broadly in agreement with $f\sigma_{8}$ observations, and show that changes to the growth of structure from ACG models are likely to be small.

\section{Discussion}
\label{discussion}

In this section, we discuss the implications of an ACG explanation to the dynamical DE preferred by observations.

\begin{figure}
    \centering
    \includegraphics[width=\columnwidth]{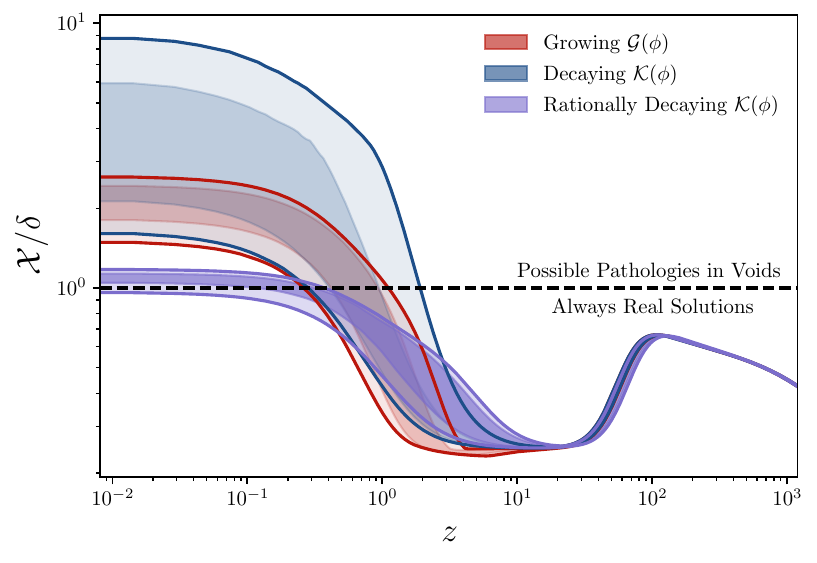}
    \caption{Vainshstein screening factor from the Growing $\mathcal{G}(\phi)$ (red), Decaying $\mathcal{K}(\phi)$ (blue) and the rationally Decaying $\mathcal{K}(\phi)$ (purple, given by Eq.~\ref{eq_rat_decay}) from constraints of the CMB, BAO and SN, with a positive ISW prior. Both Growing $\mathcal{G}(\phi)$ and Decaying $\mathcal{K}(\phi)$ violate the maximum $\mathcal{X}/\delta\le 1$, meaning they might have pathological unreal solutions in voids, see main text for more details. Different growing/decaying functions can help alleviate this, as shown for rationally decaying $\mathcal{K}(\phi)$.}
    \label{fig_chi_delta}
\end{figure}

\subsection{Non-Linear Growth}

While we leave a detailed analysis of the full linear perturbation system to future work, constraints from the CMB require the matter power spectrum at early times to remain close to that of $\Lambda$CDM. With this in mind, we have explored the linear evolution of matter perturbations in the quasistatic regime through the modified Poisson equation. The resulting growth functions are well behaved across the parameter space considered here, with the predicted $f\sigma_8$ remaining broadly consistent with $\Lambda$CDM and current measurements (see Fig.~\ref{fig_fsigma8}). We stress, however, that this does not constitute a complete test of linear stability, since the quasistatic treatment does not directly probe the exponentially growing scalar modes discussed by \citet{Cataneo2024}. Establishing their absence requires evolution of the full linear perturbation system, which we defer to future work.

For the non-linear growth of structure, implemented in the Hi-COLA code, we assume that screening is described solely by the Vainshtein mechanism. ACG models contain a $G_{3X}$ interaction that naturally gives rise to Vainshtein screening and is expected to dominate over other screening mechanisms in the region of parameter space considered here. More generally, however, ACG models may exhibit mixed screening behaviour, as described by the master screening equation of \citet{Sirera2026}. A full implementation of these additional screening mechanisms is beyond the scope of the present work and is left to future work. Under this assumption, the modified force equation is parametrised as
\begin{equation}
    F_{\mathrm{tot}} = F_{\mathrm{N}}\frac{G_{G_{4}}}{G_{\mathrm{N}}}\left(1+\beta S\right)
\end{equation}
where $F_{\mathrm{N}}$ is the standard Newtonian force experience in $\Lambda$CDM, $G_{G_{4}}/G_{\mathrm{N}}$ defines non-minimal coupling ($G_{G_{4}}/G_{\mathrm{N}}\!=\!1$ for ACG models), and $\beta$ and the Vainshstein screening term $S$ are given by
\begin{align}
    \beta &= \mu_{\mathrm{L}}-1,\\
    S &= \frac{2}{\mathcal{X}}\left(\sqrt{1+\mathcal{X}}-1\right).
\end{align}
where $\mathcal{X}$ is a theory-derived function of both redshift and the local density field, see Eq.~3.14 \citep{Wright2023}. These functions are derived in \citet{Wright2023} under the Vainshtein-screening approximation adopted here. Note that $\mathcal{X}\propto\delta$, where $\delta$ is the smoothed density contrast. An expansion at low values of $\delta$ shows that $\delta=0$ (i.e.~the mean density) implies $S=1$ and therefore returns the linear modification to the Poisson equation. For $\delta>0$, the screening factor $S$  reduces the effective strength of the fifth force, damping modifications to the Poisson equation and driving the total gravitational force towards its Newtonian value in overdense regions.

However, when $\delta <0$, there can be cases where $\mathcal{X} < -1$, leading to an imaginary $S$ \citep{Baker2018}. It remains unclear whether this is simply caused by a breakdown in approximations or is a genuine pathology of the model, but there's evidence suggesting the latter is more probable~\citep{Winther:2015pta}. To ensure this pathology is not present in a model requires that $\mathcal{X}/\delta < 1$ at all times \citep[][where $f_{\mathrm{MG}}\equiv \mathcal{X}/\delta$]{Moretti2026}. In Fig.~\ref{fig_chi_delta}, we illustrate the constraints on $\mathcal{X}/\delta$ functions from the Growing $\mathcal{G}(\phi)$ and Decaying $\mathcal{K}(\phi)$ ACG models, showing that both models follow similar paths, deviating only around redshift $3-4$, where the Growing $\mathcal{G}(\phi)$ dips below Decaying $\mathcal{K}(\phi)$ model. Rather more importantly, at low redshift ($z<1$) both models have $\mathcal{X}/\delta>1$, meaning they may exhibit pathologies in voids. This effect can be mediated by opting for a more gradual growing or decaying function in $G_{3}$ or $K$, respectively. As an example, we show in Fig.~\ref{fig_chi_delta} how this can be achieved with a rationally decaying function
\begin{equation}
    \mathcal{K}(\phi)=\frac{1}{\sqrt{1+c_{k}\phi}},
    \label{eq_rat_decay}
\end{equation}
allowing some fully-extended $\mathcal{X}/\delta <1$. 

Note that such a pathology is known to arise in pure Vainshtein theories, whereas its behaviour in the presence of mixed screening mechanisms, which may occur in ACG models, remains unclear. We therefore refer to this as a `possible' pathology in voids in Fig.~\ref{fig_chi_delta}. Determining whether this reflects a genuine inconsistency of the model or simply the breakdown of the Vainshtein-only approximation adopted here requires a more complete treatment of screening, which we leave to future work.

\subsection{Effective Neutrino Mass}

\begin{figure}
    \centering
    \includegraphics[width=\columnwidth]{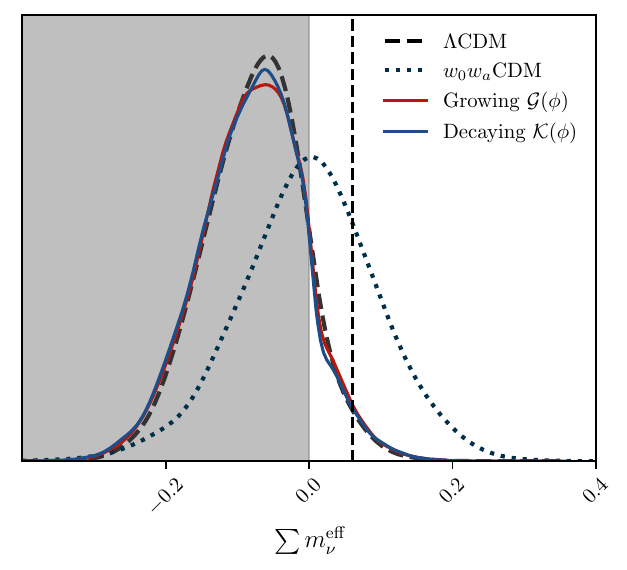}
    \caption{Marginalised posterior distributions for the effective neutrino mass in $\Lambda$CDM, dynamical DE, and the Growing $\mathcal{G}(\phi)$ and Decaying $\mathcal{K}(\phi)$ ACG models, obtained from joint CMB, BAO and SN constraints (with a positive ISW prior where applicable). The dynamical DE model peaks close to $\sum m_{\nu}^{\rm eff}\approx0$ eV, whereas the remaining models exhibit a preference for negative effective neutrino masses.}
    \label{fig_neutrino_mass}
\end{figure}

\begin{table}
    \centering
    \begin{tabular}{lcccc}
        \hline\hline
        \B & \multirow{2}{*}{$\Lambda\mathrm{CDM}$} & \multirow{2}{*}{$w_{0}w_{a}\mathrm{CDM}$} & $\mathrm{Growing}$ & $\mathrm{Decaying}$ \\
        \B & & & $\mathcal{G}(\phi)$ & $\mathcal{K}(\phi)$ \\
        \hline\hline
        \T\B $\Delta \chi_{\mathrm{MAP},\,\nu}^{2}$ & $-4.14$ & $-0.483$ & $-1.53$ & $-1.47$ \\
        \B Significance & $0.96\sigma$ & $0.513\sigma$ & $0.783\sigma$ & $0.774\sigma$ \\
        \hline
        \T $\Delta \ln Z_{\nu}$ & $0.632$ & $-1.85$ & $-0.377$ & $-1.34$ \\
        \B Betting Odds & $1.88$ & $0.158$ & $0.686$ & $0.263$ \\
        \hline
        \hline
    \end{tabular}
    \caption{We compare the improvement in $\chi^{2}$ and Bayesian evidence $Z$ for for $\Lambda$CDM, dynamical DE and the Growing $\mathcal{G}(\phi)$ and Decaying $\mathcal{K}(\phi)$ ACG models with and with neutrino masses free. We find only moderate improvements in $\chi^{2}_{\mathrm{MAP}}$ with neutrino masses free (less than $1\sigma$). Bayesian evidence is negative for all the models, with the exception of $\Lambda$CDM, showing there is a preference for a fixed neutrino mass of $\sum m_{\nu}^{\mathrm{eff}}=0.06\,\mathrm{eV}$. This shows that although the ACG models have a preference for negative neutrino mass, this preference is weak and current observations are not precise enough to distinguish the effects of massive neutrinos.}
    \label{tab_neutrino_comparison}
\end{table}

Constraints on the sum of neutrino masses have been steadily lowering as cosmological datasets become more precise, and are now approaching the minimum value required by neutrino-oscillation experiments, $\sum m_\nu \simeq 0.06,{\rm eV}$. Several studies have noted that, when modelled as an effective neutrino mass in a $\Lambda$CDM cosmology, current cosmological constraints can peak in the negative range \citep{Craig2024,Elbers2025}. This behaviour reflects the geometric role of massive neutrinos and their degeneracy with the late-time expansion history, rather than simply their suppression of structure growth \citep{LoVerde2024}. In phenomenological dynamical-DE models, the preference for negative effective neutrino mass is alleviated \citep{ElbersDESI2025}; \citet{Yang2026} showed that this is driven by the early phantom phase, while \citet{Weiner2026} interpreted the relevant effect in terms of a matter-era distance excess mediated by the high-redshift extrapolation of the dynamical-DE equation of state.

In Fig.~\ref{fig_neutrino_mass}, we show the marginal constraints on the effective neutrino mass from $\Lambda$CDM, dynamical DE and Growing $\mathcal{G}(\phi)$ and Decaying $\mathcal{K}(\phi)$ ACG models from joint constraints of the CMB, BAO and SN (and the positive ISW prior where appropriate). We see that all models, with the exception of phenomenological dynamical DE, prefer negative effective neutrino masses. This is perhaps not surprising: although the ACG models realise phantom crossing, their high-redshift evolution rapidly approaches a $\Lambda$CDM-like limit. They therefore do not significantly alter the matter-era distance interval in the way required to relax the effective neutrino-mass constraint in phenomenological dynamical-DE models \citep{Yang2026,Weiner2026}.

Using the same previously described metrics, we compare in Table~\ref{tab_neutrino_comparison} $\chi_{\mathrm{MAP}}^{2}$ and Bayesian evidence for models with and without neutrino mass as a free parameter. The difference in $\chi^{2}_{\mathrm{MAP}}$ is evaluated as
\begin{equation}
    \Delta \chi^{2}_{\mathrm{MAP},\,\nu}= \chi^{2}_{\mathrm{MAP},\,\nu\mathrm{-free}}-\chi^{2}_{\mathrm{MAP,\,fid}},
\end{equation}
where $\chi^{2}_{\mathrm{MAP},\,\nu\mathrm{-free}}$ is the $\chi^{2}$ at the MAP position with $\sum m_{\nu}^{\mathrm{eff}}$ free and $\chi^{2}_{\mathrm{MAP,\,fid}}$ the MAP with $\sum m_{\nu}^{\mathrm{eff}}=0.06\,\mathrm{eV}$. The significance is now evaluated assuming a $\chi^{2}$ distribution with 1 parameter, for neutrino mass, instead of the 2 used in the previous model comparison. Similarly, the Bayesian evidence is given by
\begin{equation}
    \Delta \ln Z_{\nu} = \ln Z_{\nu\mathrm{,\,free}}-\ln Z_{\nu\mathrm{,\,fid}},
\end{equation}
where $\ln Z_{\nu\mathrm{,\,free}}$ has $\sum m_{\nu}^{\mathrm{eff}}$ free and $\ln Z_{\nu\mathrm{,\,fid}}$ with $\sum m_{\nu}^{\mathrm{eff}}=0.06\,\mathrm{eV}$. 

Although ACG models prefer a negative neutrino mass, the improvement in $\chi^{2}_{\mathrm{MAP}}$ is relatively low (less than $1\sigma$ in all cases) and, with the exception of $\Lambda$CDM, Bayesian evidence prefers the models with neutrino mass set to $\sum m_{\nu}^{\mathrm{eff}}=0.06\,\mathrm{eV}$. This shows that the preference for negative neutrino masses is only moderately significant for $\Lambda$CDM, while for the ACG models, this is of low significance, and the expansion history constraints are currently not precise enough to distinguish the dominant ACG behaviour from the sub-dominant effects of massive neutrinos. Hence the negative neutrino mass tension remains in this model, but for the time being the significance of this tension is low.

\begin{figure*}
    \centering
    \includegraphics[width=0.9\textwidth]{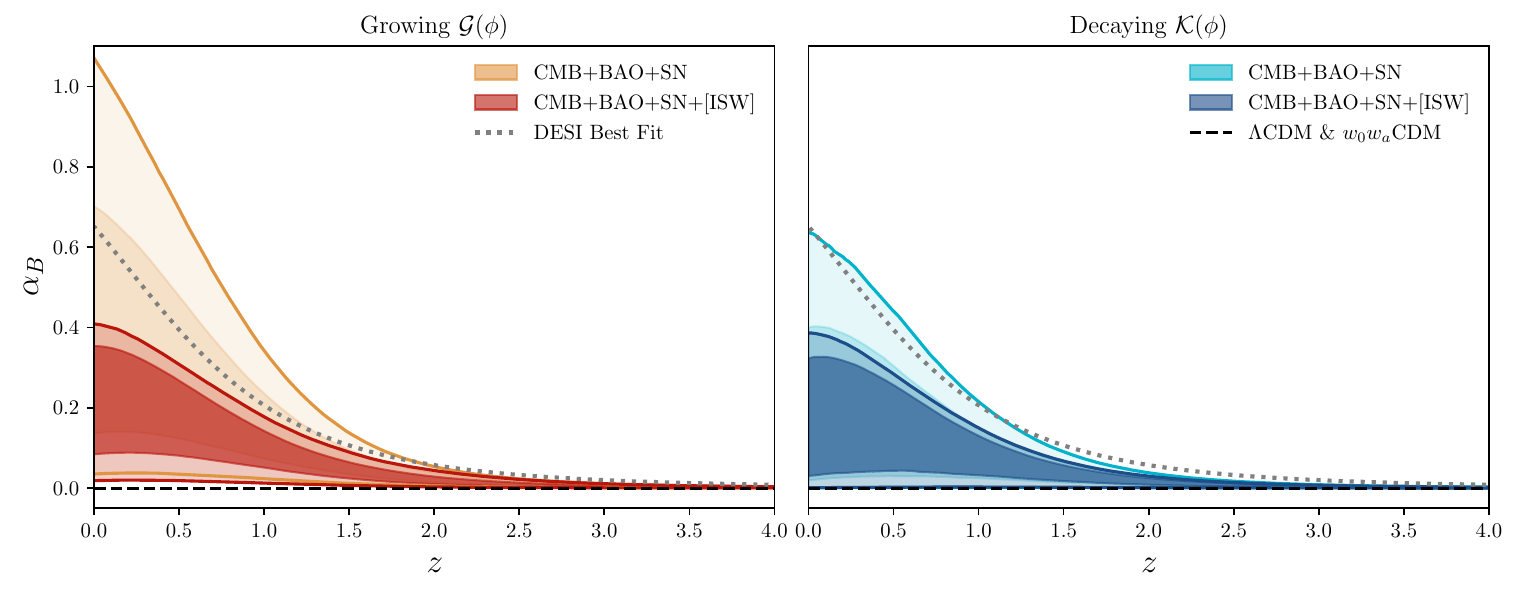}
    \caption{Constraints on the ACG $\alpha_{B}$ functions from measurements of the CMB, BAO and SN, shown in comparison to the best-fit function from DESI \citep[using DESI fullshape and BAO measurements, DES year 5 SN and Planck CMB;][]{Ishak2025}. This figure follows the same form as Fig.~\ref{fig_BAO}.  The DESI best fit profile uses an effective field theory description of DE resulting in a function that is larger than what is predicted from the ACG models, in particular with the additional constraint from the positive ISW prior. Since the parameterisation employed in \citet{Ishak2025} does not include ISW constraints it is likely that part of the parameter space, including the best-fit values, produce negative ISW. }
    \label{fig_alpha_B}
\end{figure*}

\subsection{Braiding Strength}

In Fig.~\ref{fig_alpha_B}, we compare the predictions for $\alpha_{B}$ (defined in Appendix~\ref{app_alphas}) for the Growing $\mathcal{G}(\phi)$ and Decaying $\mathcal{K}(\phi)$ from constraints of the CMB, BAO and SN in comparison to best-fit predictions from \citet{Ishak2025} using an effective field theory description of DE from constraints of the CMB, DES year 5 SN and DESI fullshape and BAO measurements. Predictions of ACG models generally point towards lower values than the best-fit measurements from DESI. This discrepancy is larger once we impose the positive ISW prior. Since \citet{Ishak2025} do not include constraints on the ISW, it is likely that a significant portion of the posterior, including the best-fit constraint, result in negative ISW cross-correlations.

Since our ACG models have $G_4(\phi)=1/2$ , the Planck mass run rate parameter $\alpha_M$, also constrained in \citet{Ishak2025}, is automatically zero.

\section{Conclusions}
\label{conclusions}

In this paper, we show that the dynamical DE model preferred by observations, a model in which DE is phantom at early times and crosses the phantom divide at late times, can be replicated by a Horndeski gravity model based on the Cubic Galileon model with broken shift symmetry. Previous Lagrangian-level constructions have pursued this behaviour either through non-minimal coupling \citep{Ye2025,Wolf2025,Wolf2025b} or, within minimally coupled kinetic-braiding models, by breaking shift symmetry with an additive $\phi$-dependent potential \citep{Tsujikawa2025,Wolf2026,Calderon2026}. Complementary stable-basis explorations have also shown that viable minimally coupled Kinetic Gravity Braiding (KGB) models can realise phantom crossing \citep{Cataneo2026}. Here, we instead provide an explicit minimally coupled Lagrangian realisation close to the Cubic Galileon model, in which phantom crossing is obtained by multiplying either the kinetic term by a decaying $\mathcal{K}(\phi)$ function or the kinetic-braiding term by a growing $\mathcal{G}(\phi)$ function. Our models also break shift symmetry, but by design only do so at late cosmological times. While we consider an exponentially decaying $\mathcal{K}(\phi)$ and a linearly growing $\mathcal{G}(\phi)$ function in this paper, phantom crossing can be realised more generally in models where the kinetic-to-braiding strength $\mathcal{K}/\mathcal{G}^{2/3}$ decreases \citep{Hallam2026}. Throughout this work, we additionally require all viable models to satisfy the no-ghost and gradient-stability conditions, demonstrating that the phantom crossing is achieved without pathological scalar perturbations.

We show the implications of requiring consistency with a positive galaxy-ISW cross-correlation for this Horndeski-based solution to phantom crossing. We find that this requirement pushes the constraints towards an EoS that is shallower (see Fig.~\ref{fig_EoS}), leading to modifications to the Poisson and lensing potentials that are more gradual (see Fig.~\ref{fig_mu_sigma}). This leads to growth of structure that is not significantly different from $\Lambda$CDM, but with a slight preference for greater growth at very late times ($z<0.25$). Although the positive-ISW prior removes a substantial region of the ACG parameter space, often shifting the constraints away from the best-fit regions, we find that the predictions for BAO and distance modulus are not significantly altered. More generally, the requirement of a positive ISW cross-correlation appears to favour ACG models with gradual evolution in both the DE EoS and the modified gravity functions. Such models not only remain compatible with the observed sign of the ISW cross-correlation, but also tend to alleviate the Vainshtein-void pathologies found in void environments \citep{Barreira2015, Winther:2015pta, Moretti2026} (see also Sirera et al. in preparation, for an analysis of void pathologies in a combined chameleon–Vainshtein scenario), suggesting that viable ACG models occupy a restricted region of parameter space characterised by relatively shallow phantom behaviour. We stress, however, that the ISW treatment adopted here uses only the sign of the cross-correlation and does not constitute a full statistical analysis of existing galaxy-ISW measurements.

We use several model comparison metrics to quantify how the Horndeski ACG models compare to dynamical DE. We find that both dynamical DE and the ACG models provide better fits to the data in comparison to $\Lambda$CDM, with an improvement of $\Delta\chi^{2}_{\mathrm{MAP}}\simeq-11$, but with dynamical DE generally providing better fits $\Delta\chi^{2}_{\mathrm{MAP}}\simeq-13$. In terms of Bayesian evidence, we find the ACG models provide slightly better evidence ratios. In either case, the $\chi^{2}_{\mathrm{MAP}}$ or Bayesian evidence $\ln Z$, show a moderate preference for dynamical DE and the ACG models over $\Lambda$CDM but not at definitive level. The advantage of the ACG models is they provide a viable theoretical formulation, at the Lagrangian level, to the dynamical DE behaviour preferred by observations.

We also consider the implications of Vainshtein screening within voids. While pathological screening factors have been identified in pure Vainshtein theories \citep{Baker2018,Moretti2026}, it remains unclear whether this behaviour persists in ACG models, which may exhibit multiple screening mechanisms \citep{Sirera2026}. Nevertheless, under the Vainshtein-only approximation adopted in this work, our models can exhibit regions with pathological screening factors. This behaviour can be partially alleviated by adopting more gradual forms for $\mathcal{K}(\phi)$ or $\mathcal{G}(\phi)$, as discussed further in \citet{Hallam2026}. Both the positive ISW prior and the requirement to reduce these void pathologies favour more gradual functional forms, leading to phantom crossing occurring slightly earlier, around $z\sim0.75$. Should viable models continue to exhibit regions where $\mathcal{X}/\delta>1$, a more complete treatment of screening will be required to determine whether this reflects a genuine theoretical inconsistency or simply the breakdown of the Vainshtein-only approximation.

Lastly, we investigate whether these models also alleviate the preference for negative neutrino mass found in $\Lambda$CDM \citep{Craig2024,Elbers2025,ElbersDESI2025}. In dynamical DE, this alleviation has been attributed to the early phantom phase, which shifts the effective neutrino mass posterior towards $\sum m_{\nu}^{\mathrm{eff}}=0$ \citep{Yang2026}. We show that the Horndeski ACG models do not resolve this tension, but instead prefer negative neutrino mass. This is because the ACG models recover $\Lambda$CDM-like behaviour at high redshift and only develop a relatively shallow, late-time phantom phase. They therefore do not generate the matter-era distance-excess behaviour that allows phenomenological dynamical-DE models to relax the effective neutrino-mass constraint \citep{Yang2026,Weiner2026}. However, using Bayesian evidence, we show that the preference for a negative neutrino mass is only moderately significant in $\Lambda$CDM, while for dynamical DE and the ACG models the data is not sufficiently constraining for neutrino masses to be free, and keeping neutrino mass fixed to $\sum m_{\nu}^{\mathrm{eff}}=0.06\,\mathrm{eV}$ is actually preferred. This suggests that current data are unable to robustly disentangle the effects of neutrino mass from the dominant background evolution of the ACG models.

In future work, we plan to study the linear perturbation theory constraints of the ACG models, by integrating the 
\texttt{Hi-COLA} simulation software for Horndeski gravity with a Boltzmann solver such as \texttt{hi\_class} \citep{hiclass2017} or \texttt{EFT-CAMB} \citep{Hu2014}. This will enable us to use the full Planck likelihood and two point statistics from DES\footnote{\href{https://www.darkenergysurvey.org/}{https://www.darkenergysurvey.org/}} \citep{DES2026}, Baryon Osciliation Spectroscopic Survey \citep[BOSS;][]{Alam2017} and DESI. We will also use \texttt{Hi-COLA} to study the evolution of these models in the non-linear regime, enabling the prediction of large-scale structure observations from DESI and weak lensing surveys such as \textit{Euclid}\footnote{\href{https://www.euclid-ec.org/}{https://www.euclid-ec.org/}} \citep{Euclid2025}, allowing them to be validated or ruled out by Stage IV survey data.

\section*{Acknowledgements}

KN, TB and SS acknowledge support from Royal Society grant number URF\textbackslash R\textbackslash 231006 and ERC Starting Grant SHADE (grant no.\,StG 949572). JH is supported by a UKRI/STFC PhD studentship. Numerical computations were carried out on the SCIAMA High Performance Compute (HPC) cluster which is supported by the ICG and the University of Portsmouth.

\section*{Data Availability}

The data underlying this paper are derived from publicly available observational datasets. Details of these datasets are provided in the main text and/or references. All software and data-processing pipelines developed for this work are publicly available.



\bibliographystyle{mnras}
\bibliography{biblio} 



\appendix

\section{Effective Field Theory and Stability Conditions}
\label{app_alphas}

In this appendix, we write out the effective field theory alpha parameters and stability conditions for the luminal set of Horndeski gravity, following \citet{Bellini2014}, but rewritten with derivatives in $\ln a$.

\subsection{Effective Field Theory Parameterisation}

Horndeski gravity is often expressed in terms of $\alpha$ parameters, defined as a reparameterisation of the effective field theory (EFT) of DE \citep{Bellini2014}. These are defined in terms of the Horndeski free functions $K$, $G_{3}$ and $G_{4}$ as
\begin{equation}
\begin{split}
    \alpha_{M}=&\frac{\phi^{\prime}G_{4\phi}}{G_{4}},\\
    \alpha_{B}=&\frac{\phi^{\prime}}{G_{4}}\left(X G_{3X}-G_{4\phi}\right),\\
    \alpha_{K}=&\frac{1}{2E^{2}G_{4}}\bigg[2X\bigg(K_{X}+2XK_{XX}-2G_{3\phi}-2XG_{3\phi X}\bigg) \\ &+ 12E^{2}\phi^{\prime}X\left(G_{3X} + XG_{3XX}\right)\bigg],
\end{split}
\end{equation}
which is expressed in the luminal Horndeski framework, implying $\alpha_{T}=0$.

\subsection{Stability Conditions}
\label{subsec:aint_afraid_of_no_ghosts}

The stability of a given Horndeski model requires
\begin{equation}
    Q_{s}=\frac{4M_{p}^{2}G_{4}D}{(2-\alpha_{B}^{2})}>0,\quad D\equiv\alpha_{K}+\frac{3}{2}\alpha_{B}^{2},
\end{equation}
\begin{equation}
\begin{split}
    c_{s}^{2}=-\frac{1}{D}\Bigg\{(2-\alpha_{B})\bigg[\frac{E^{\prime}}{E}&-\frac{1}{2}\alpha_{B}-\alpha_{M}\bigg]-\alpha_{B}^{\prime}\\
    &+\frac{3}{2G_{4}}\sum \frac{\rho_{i}}{E^{2}}(1+w_{i})\Bigg\}>0,
\end{split}
\end{equation}
which ensures the model has no ghost ($Q_{s}>0$) and gradient ($c_{s}^{2}>0$) instabilities.

\section{The Numerical Solver}
\label{solver}

The public release of \texttt{Hi-COLA}, corresponding to the implementation presented by \citet{Wright2023}, is available from github.\footnote{\url{https://github.com/Hi-COLACode/Hi-COLA}} The numerical framework described in this appendix corresponds to \texttt{Hi-COLA} version 2 (v2), which incorporates the general Horndeski implementation and solver improvements presented in this work. We intend to make this version publicly available through the same GitHub repository upon acceptance of this paper.

This appendix describes the numerical implementation of \texttt{Hi-COLA} v2, including its framework for computing the cosmological expansion history and scalar-field evolution for general Horndeski models.

We then constructed the \texttt{HorndeskiModel} class, that inherits the base StandardModel class, but is tasked with solving the expansion history and scalar field evolution within the Horndeski framework outlined in Sec.~\ref{sec_Horndeski_theory}. The class is designed to take in any user-defined Horndeski model, where the user need only define the Horndeski free functions $K$, $G_{3}$ and $G_{4}$. Alternatively, the user may choose to use predefined Horndeski models. These can be used by instead calling the classes \texttt{CubicGalileon} for CG, \texttt{ESS} for the Extended Shift-Symmetric model \citep[see][]{Wright2023} or \texttt{AsymCubicGalileon} for the ACG models.

The class then constructs the model symbolically using the \texttt{sympy} python module \citep{Sympy2017}, together with the Friedmann equations and scalar field functions are all constructed symbolically following Sec.~\ref{sec_Horndeski_theory}. These are then converted into callable python functions. A benefit of a class based routine is these functions need only be defined once, increasing the efficiency of the software particularly when many parameter calls are required.

All likelihood functions and sampling algorithms are implemented within \texttt{Hi-COLA} through the \texttt{Sampler} class. This class can either be instantiated directly by the user or accessed via the \texttt{HiCOLA.scripts.run\_sampler} script, which reads a \texttt{yaml} configuration file and supports parallel execution through Python's \texttt{multiprocessing} module.

\subsection{Explicit Form of the Background Equations}
\label{app_background}

For completeness, we provide the auxiliary functions entering Eqs.~\ref{eq:E_prime} and \ref{eq:phi_primeprime}. Following \citet{Wright2023}, the Raychaudhuri equation may be written as Eq.~\ref{eq:E_prime} with
\begin{equation}
\begin{split}
    \mathcal{A} = & A\bigg(-\frac{K}{2E^{2}}+\frac{XG_{3\phi}}{E^{2}}-2\phi^{\prime}G_{4\phi}-\frac{2XG_{4\phi\phi}}{E^{2}}\\
    &-\frac{3}{2E^{2}}\sum_{i}\hat{P}_{i}-3G_{4}\bigg)-B_{2}\left(XG_{3X}-G_{4\phi}\right),
\end{split}
\end{equation}
and 
\begin{equation}
    \mathcal{B} = 2AG_{4}+B_{1}\big(XG_{3X} - G_{4\phi}\big),
\end{equation}
where $\hat{P}_{i} = w_{i} \hat{\rho}_{i}$ is the pressure of constituent $i$.

The scalar-field equation of motion is given by eq.~\ref{eq:phi_primeprime} where
\begin{equation}
    \begin{split}
        A = &K_{X} - 2G_{3\phi} +2XG_{3\phi X} + E^{2}\bigg[ 6\phi^{\prime}\big(G_{3X} + XG_{3XX}\big)\\
        &+\phi^{\prime 2}\big(K_{XX}-2G_{3\phi X}\big)\bigg],
    \end{split}
\end{equation}
\begin{equation}
    B_{1} =6XG_{3X}-6G_{4\phi},
\end{equation}
and
\begin{equation}
\begin{split}
    B_{2} = & 3\phi^{\prime}\left(K_{X}-2G_{3\phi}+2XG_{3X\phi}\right) + \phi^{\prime 2}\left(K_{X\phi}-2G_{3\phi\phi}\right)\\
    &-\frac{K_{\phi}}{E^{2}}-12G_{4\phi}+18XG_{3X}+\frac{2XG_{3\phi\phi}}{E^{2}}.
\end{split}
\end{equation}

\subsection{Initial Conditions}
\label{ICs}

The evolution of the expansion history and scalar field is governed by the closure relation (Eq.~\ref{eq_hubble_expansion}), the second Friedmann equation (Eq.~\ref{eq:E_prime}) and the equation of motion (Eq.~\ref{eq:phi_primeprime}). In Horndeski gravity, these functions are coupled ODEs that must be solved numerically for the expansion rate $E$, the scalar field $\phi$, and its derivative $\phi^{\prime}$. However, before we can begin solving the coupled ODEs, we must first define their starting values. The solver can, in principle, start now ($z=0$) and work backwards, as carried out in the initial \texttt{Hi-COLA} code, or it can start at very high redshift ($z\approx1000$) and then evolve forwards.

In the backward solver, the normalised expansion rate becomes very trivial to define: we simply set $E=1$; but the scalar field $\phi$ and its derivative $\phi^{\prime}$ are difficult to set a priori. For a shift-symmetric model, there is no $\phi$-dependence, so $\phi$ can be set to an arbitrary value. We then solve the closure relation for $\phi^{\prime}$, which we carry out symbolically using \texttt{sympy}. For general $\phi$-dependent models, we must either set the value of $\phi$ or $\phi^{\prime}$ and then use the closure relation to solve for the other term. However, at late times, this tends to be numerically unstable due to the scalar field being large and very sensitive to the initial chosen values. 

A forward solver resolves some of these issues, since we can assume that at very early times, the scalar field is relatively small and the effects of DE are negligible. This means we can choose to set the scalar field to a very small value, such as $\phi_{\mathrm{ini}}=10^{-5}$. However, unlike the backwards solver, $E$ has no natural starting value. To set $E$, we can take advantage of the fact that, in the early Universe, DE is negligible and the dynamics are dominated by radiation and matter. In other words, at early times we can assume the Universe is very close to $\Lambda$CDM, meaning we can set the initial value of $E=E_{\Lambda \mathrm{CDM}}(z_{\mathrm{ini}})$ to its $\Lambda$CDM value at an initial redshift $z_{\mathrm{ini}}$. For a shift-symmetric model, we then solve for $\phi^{\prime}$. For $\phi$-dependent models, we can either set $\phi$ or $\phi^{\prime}$ to be very small (of the order of $10^{-5}$) or we can solve them jointly using both the closure relation and the second Friedmann equation by setting $E^{\prime}=E_{\Lambda \mathrm{CDM}}^{\prime}(z_{\mathrm{ini}})$ to its $\Lambda$CDM value. This allows us to reproduce very stable expansion functions but comes at the cost of producing $\phi$ and $\phi^{\prime}$ functions that are numerically unstable, since at this time both $\phi$ and $\phi^{\prime}$ are very small and sensitive to numerical noise. In practice, we find it best to set $\phi\simeq10^{-5}$ and solve the closure relation for $\phi^{\prime}$.

Solving the closure relation for $\phi$ or $\phi^{\prime}$ may result in more than one solution. \texttt{Hi-COLA} will evolve the expansion rate and scalar field for all real-valued solutions. However, in practice, most models (including CG, ACG and ESS) produce only a single real-root, with the rest of the solutions being imaginary. In the rare instances where multiple real roots are present, we generally find only a single solution remains viable with observations, while the other solutions require extreme shifts to the expansion rate and can therefore be ignored.

\subsection{Ordinary Differential Equation Solver with Constraint}

Once the initial conditions have been set, we now evolve $E$, $\phi$ and $\phi^{\prime}$ with the second Friedmann equation (Eq.~\ref{eq:E_prime}) and the equation of motion (Eq.~\ref{eq:phi_primeprime}) using the \texttt{solve\_ivp} ODE solver from \texttt{scipy}. However, out treatment of $E$ is slightly different to the original implementation presented in \citep{Wright2023}. Unlike $\phi$ and $\phi^{\prime}$, $E$ is not a dynamical function, but instead a constrained quantity set by the closure relation. So while Eq.~\ref{eq:E_prime} provides the derivative for $E$, the value of $E$ obtained after a small integration step will not necessarily satisfy the closure relation. This violation remains small in the original \texttt{Hi-COLA} implementation, but becomes more significant for $\phi$-dependent models, such as the ones considered in this paper. To ensure the closure relation is satisfied, we use the new $E$ as an initial guess and employ a Newton root solver function (specifically \texttt{newton} from \texttt{scipy}) to correct $E$ for the closure relation. This ensures $E$ does not drift and thus provide invalid scalar field solutions. To ensure stability, we require that the root finding for $E$ has a relative tolerance of $10^{-5}$ and that the ODE solver has absolute and relative tolerances of $10^{-9}$. In either case, these tolerance values can be specified by the user.

\subsection{Scaling Relations}
\label{subsec:scaling_relations}

In solving the ODE forwards from high redshift, we find the raw outputs for $E$ may result in $E(z=0)\neq1$, since we deviate away from the assumed $\Lambda$CDM initial conditions. This translates to a mismatch between the input and output Hubble constant, which we can evaluate as
\begin{equation}
    H_{0}= H_{0}^{\mathrm{GR}}\,\hat{E}(z=0),
\end{equation}
where we define hat terms, such as $\hat{E}$, as raw outputs from the numerical ODE solver. To correct $\hat{E}$ so that $E(z=0)=1$, as is required by definition, we define the the following ratio
\begin{equation}
    f_{H}=\frac{H_{0}}{H_{0}^{\mathrm{GR}}}
\end{equation}
and apply the following corrections to the raw outputs
\begin{equation}
\begin{split}
    &E=\frac{\hat{E}}{f_{H}},\quad \rho_{i}=\frac{\hat{\rho}_{i}}{f_{H}^{2}},\\
    &\phi=f_{H}\hat{\phi},\quad \phi^{\prime}=f_{H}\hat{\phi}^{\prime},\quad\text{and}\quad \phi^{\prime\prime}=f_{H}\hat{\phi}^{\prime\prime},
\end{split}
\end{equation}
to correctly scale the raw outputs self-consistently.

\subsection{Mass Scales}
\label{subsec:mass_scales}

In Hi-COLA, the scalar field and the three Horndeski functions are normalised to absorb the appropriate mass scales. We denote the original dimensionful quantities appearing in the Horndeski Lagrangian (Eq.~\ref{eq:luminal_horndeski_action}) with a tilde, while the corresponding dimensionless variables used throughout Hi-COLA are written without a tilde. The scalar field is defined as
\begin{equation}
    \tilde{\phi}=M_{s}\phi,
\end{equation}
where $M_{s}$ is its associated mass. The three functions are defined as
\begin{equation}
    \tilde{K}=M_{K}^{4}K,\quad \tilde{G}_{3}=M_{G3}G_{3},\quad\text{and}\quad\tilde{G}_{4}=M_{G4}^{2}G_{4},
\end{equation}
where $M_{K}$, $M_{G3}$ and $M_{G4}$ are the mass scales associated with each function. In Hi-COLA, we define the following ratios
\begin{equation}
    \begin{split}
        &R_{sp}=\frac{M_{s}}{M_{p}},\quad R_{Kp}=\frac{M_{K}^{4}}{H_{0}^{2}M_{p}^{2}},\quad R_{G3p}=\frac{M_{G3}}{M_{p}},\quad\text{and}\\
        &R_{G4p}=\frac{M_{G4}^{2}}{M_{p}^{2}}.
    \end{split}
\end{equation}
These ratios can be specified but are assumed, by default, to all be equivalent to unity.

\begin{figure*} 
    \centering
    \includegraphics[width=0.8\textwidth]{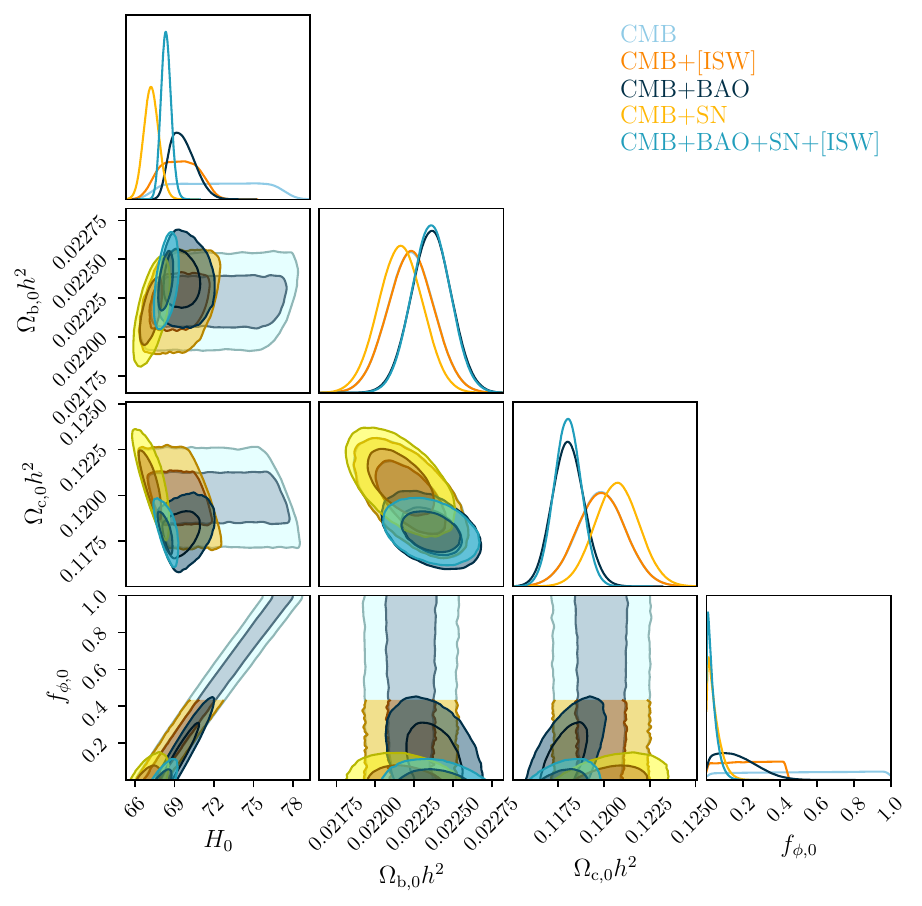}
    \caption{Constraints on the CG model from the CMB, BAO, SN and a positive ISW prior. CMB only measurements allow large $H_{0}$ values, sufficient to resolve the Hubble tension; the discrepancy between local measurements \citep{Riess2022} and early Universe constraints \citep{Planck2020}. However, with the inclusion of other datasets and the positive ISW prior, the constraints are pushed towards $f_{\phi,0}\rightarrow 0$, consistent with $\Lambda$CDM. This is likely driven by the preference for phantom crossing in the data, that is not possible in CG models.}
    \label{fig_constraints_CG}
\end{figure*}

\begin{figure*} 
    \centering
    \includegraphics[width=0.9\textwidth]{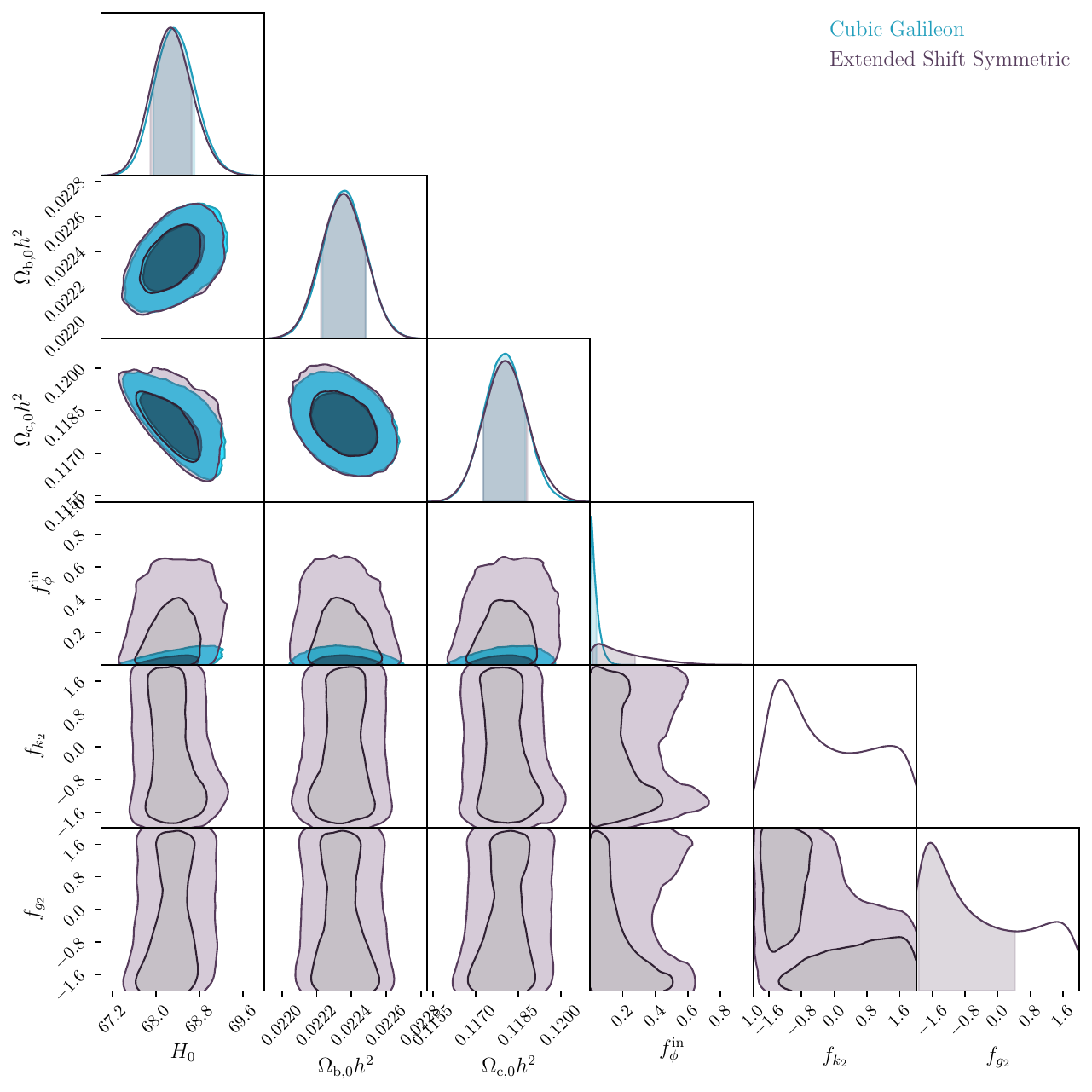}
    \caption{Constraints on the extended Shift Symmetric model with respect to the CG model from joint fits to CMB, BAO, SN and a positive ISW prior, showing no significant constraints on the ESS parameters $f_{k_{2}}$ and $f_{g_{2}}$, despite wider constraints on $f_{\phi}^{\mathrm{in}}$. There is a slight preference for $f_{k_{2}}\neq0$ and $f_{g_{2}}\neq0$, i.e.~away from the CG limit, caused by a reduction in the phantom EoS in these parts of ESS parameter space, albeit without crossing the phantom divide.}
    \label{fig_constraints_ESS}
\end{figure*}

\section{Parameter Constraints on Shift Symmetric Models}
\label{app_shift_symmetric}

Due to the simpler dynamics of the models, shift symmetric Horndeski models have been studied in great detail, including in the context of \texttt{Hi-COLA}, where simulations were first developed for shift symmetric models \citep{Wright2023}. One of the simplest Horndeski models is the CG model, where the functional form of the model is defined in Sec.~\ref{sec:cubic_galileon}. In Fig.~\ref{fig_constraints_CG}, we show the parameter constraints of the CG model from various data sets. With constraints from the CMB only, we see a complete degeneracy in the $f_{\phi,0}$ term, which is broken when adding constraints from either BAO, SN or a positive ISW prior. The joint constraints from the CMB, BAO and SN push the constraints towards $f_{\phi,0}\rightarrow0$, effectively pushing the CG model towards $\Lambda$CDM. This is perhaps unsurprising given the BAO and SN show a preference for phantom crossing, while the CG model is always phantom. In fact, stable phantom crossing is strongly constrained in shift-symmetric Horndeski models \citep{Traykova2021,Tsujikawa2025,Linder2025}, and purely shift-symmetric models without a potential do not realise the desired crossing behaviour \citep{Calderon2026}.

We constrain the Extended Shift Symmetric (ESS) model \citep{Wright2023} to test whether a more complex shift symmetric model could explain the data without phantom crossing. The ESS model is defined with the following functional form
\begin{align}
    &K(X) = k_{1}X+k_{2}X^{2},\\
    &G_{3}(X)=g_{31}X+g_{32}X^{2},\\
    &G_{4}=\frac{1}{2},
\end{align}
where we can use the Tracker ansatz to define the constants, as was previously done for the CG model,
\begin{align}
    &k_{1} = -\frac{12f_{\phi}\Omega_{\mathrm{DE},0}}{2+f_{k_{2}}}\quad\mathrm{and}\quad g_{31} = \frac{4f_{\phi}\Omega_{\mathrm{DE},0}}{2+f_{g_{2}}},
\end{align}
where $f_{k_{2}}$ and $f_{g_{2}}$ are additional free parameters of the ESS model that define the constants
\begin{equation}
    k_{2}=f_{k_{2}}k_{1}\quad\mathrm{and}\quad g_{32}=f_{g_{2}}g_{31}.
\end{equation}
In Fig.\ref{fig_constraints_ESS}, we show the joint constraints from the CMB, BAO and SN in comparison to the CG. Constraints on the shared parameters are generally very similar with the exception of $f_{\phi,0}$ which is now much broader. However, we find no significant constraints on the new parameters $f_{k_{2}}$ and $f_{g_{2}}$, which remain consistent with the CG limit $f_{k_{2}}=0$ and $f_{g_{2}}=0$, generally filling the prior range $[-2,2]$ with weak bimodal peaks towards negative $f_{k_{2}}$ or $f_{g_{2}}$. These peaks are driven by regions of parameter space with a lower DE equation of state, moving closer to $w_{\rm DE}=-1$, but their significance is insufficient to justify the additional model complexity. Overall, the ESS model remains consistent with the CG model, and hence with the $\Lambda$CDM limit, providing no evidence for the phantom-crossing behaviour preferred by current observations.


\bsp	
\label{lastpage}
\end{document}